\documentclass[useAMS,usenatbib]{mn2e}
\usepackage[utf8]{inputenc}
\usepackage{graphicx}
\usepackage{verbatim}
\usepackage{booktabs}
\usepackage{txfonts}
\usepackage{gensymb}
\usepackage[Symbol]{upgreek}
\usepackage{hyperref}
\usepackage{xcolor}
\usepackage{times}
\usepackage{natbib}
\usepackage{color}
\usepackage[all]{hypcap} 
\usepackage{ulem}
\usepackage{mathptmx}   
\usepackage{latexsym}
\usepackage{amssymb}

\newcommand{\apjs}{ApJS}

\newcommand{\apjl}{ApJL}

\newcommand{\aap}{A \& A}

\newcommand{\enzo}{\it{\small ENZO}}

\usepackage[colorinlistoftodos]{todonotes}
 \begin{document}
 
\title[Cosmological simulations of  primordial magnetic fields from CMB constraints] {Simulations and observational tests of primordial magnetic fields from Cosmic Microwave Background constraints }
\author[F. Vazza,  D. Paoletti, S.Banfi,  F. Finelli, C. Gheller, S. O'Sullivan, M. Br\"{u}ggen]{F. Vazza$^{1,2,3}$\thanks{E-mail: franco.vazza2@unibo.it}, D. Paoletti$^{4,5}$,  S. Banfi$^{1,3}$, F. Finelli$^{4,5}$,  C. Gheller $^{3}$,
\newauthor
S.~P.~O'Sullivan$^{6}$, M. Br\"{u}ggen$^{2}$
\\
$^{1}$ Dipartimento di Fisica e Astronomia, Universit\'{a} di Bologna, Via Gobetti 92/3, 40121, Bologna, Italy\\
$^{2}$ Hamburger Sternwarte, Gojenbergsweg 112, 21029 Hamburg, Germany\\
$^{3}$INAF, Istituto di Radio Astronomia, Via Gobetti 101, 40129 Bologna, Italy\\
$^{4}$INAF, Osservatorio di astrofisica e scienza dello spazio, INAF, Via Gobetti 101, 40129 Bologna, Italy\\
$^{5}$INFN, Sezione di Bologna, Via Irnerio 46, 40126 Bologna, Italy\\
$^{6}$School of Physical Sciences and Centre for Astrophysics \& Relativity, Dublin City University, Glasnevin, D09 W6Y4, Ireland\\
}

\date{Received / Accepted}
\maketitle
\begin{abstract}
We present the first cosmological simulations of primordial magnetic fields derived from the constraints by the Cosmic Microwave Background observations, based on the fields' gravitational effect on cosmological perturbations. 
We evolved different primordial magnetic field models with the {\enzo} code and compared their observable signatures (and relative differences) in galaxy clusters, filaments and voids. 
The differences in synchrotron radio powers and Faraday Rotation measure from galaxy clusters are generally too small to be detected, whereas differences present in filaments will be testable with the higher sensitivity of the Square Kilometre Array. However, several statistical full-sky analyses, such as the cross-correlation between galaxies and diffuse synchrotron power, the Faraday Rotation structure functions from background radio galaxies, or the analysis of arrival direction of Ultra-High-Energy Cosmic Rays, can already be used to constrain these primordial field models.\\

\end{abstract}

\label{firstpage} 

\section{Introduction}
\label{sec:intro}

Although we understand by and large how the inhomogeneities of the cosmic microwave background (CMB) at $z \sim 1100$ are related to the distribution of dark (DM) and baryonic matter, the 
 origin of extragalactic magnetic fields is still a puzzle \citep[e.g.][]{2002RvMP...74..775W, 2012SSRv..166....1R,wi11}. The magnetic fields observed in galaxies \citep[e.g.][for recent work on the subject]{beck12,2014ApJ...783L..20P,2016MNRAS.457.1722R,2017arXiv170405845R,2019Galax...7...45S,2020MNRAS.497..336S}, or in the intracluster medium \citep[e.g.][]{do99,br05,ry08,xu09,bm16}, 
 are likely to be the result of amplification of weak seed fields \citep[e.g.][for recent reviews]{review_dynamo}. However, it remains unclear whether such seeds were already present at the epoch of the CMB, or whether they arose during  galaxy formation, through magnetised winds and jets. 
 
Several mechanisms to generate primordial seed fields have been suggested. These may either involve inflation or take place in the post-inflationary epoch, with the latter referred historically to as causal generation mechanisms.  Causal magnetic fields  can be possibly generated with large amplitudes, but suffer from small coherence lengths (with the size of the Hubble radius at the generation time being the maximum) and they thus require inverse cascade mechanisms to transfer energy to the largest scales. Among causal generation mechanisms, those associated with phase transitions as the electroweak or the QCD \citep[e.g.][]{Quashnock:1988vs,Vachaspati:1991nm,Baym:1995fk,Sigl:1996dm,Hindmarsh:1997tj,Grasso:1997nx,Ahonen:1997wh,Boyanovsky:2005ut,Caprini:2009yp,Tevzadze:2012kk,Zhang:2019vsb,Ellis:2019tjf} are extremely important, but they require a first-order phase transition, which is currently disfavoured. Alternative causal mechanisms involve second-order perturbations via the Harrison mechanism and recombination, but these can generate only very weak final fields \citep{Fidler:2015kkt,Fenu:2010kh,Matarrese:2004kq}. A common characteristics of causal generation mechanisms is that the magnetic fields they produce  have a scale-dependence, with a spectral index equal or greater than 2 \citep{Durrer:2003ja}. 

On the contrary, inflation can generate magnetic fields with different coherence lengths and scale dependencies (however, in this work we assume $-2.9$ as minimal index to avoid infrared divergences in the energy-momentum tensor of the fields). Magnetic fields  can be generated during inflation by breaking the conformal invariance for the electromagnetic field, or by coupling it to other light fields   \citep[e.g.][]{1988PhRvD..37.2743T,Ratra:1991bn,Giovannini:2000dj,Tornkvist:2000js,Bamba:2003av,Ashoorioon:2004rs,Demozzi:2009fu,Kanno:2009ei,Caldwell:2011ra,Jain:2012jy,Fujita:2015iga}, and depending on the specific mechanism the resulting fields have different characteristics.
Such primordial seed fields are found to produce either small \citep[$\leq ~\rm Mpc$, e.g.][]{1967SvA....10..634C}, or large \citep[e.g.][]{1970SvA....13..608Z,1988PhRvD..37.2743T} coherence lengths, which may still persist today \citep[e.g.][]{2018CQGra..35o4001H}, at least in the  emptiest cosmic regions. Primordial magnetic fields can also possibly carry information on the generation of primordial helicity \citep[e.g.][]{2005A&A...433L..53S,2009IJMPD..18.1395C,2016PhyS...91j4008K}. The subsequent amplification of these seed fields, plausibly by the dynamo mechanism \citep[see][for recent reviews]{2016JPlPh..82f5301F,review_dynamo}, further adds to the theoretical predictions of the primordial generation mechanisms.

Several different observations can be used to constrain primordial magnetic fields generated prior to Big Bang Nucleosynthesis.
Magnetic fields with a primordial origin, modelled as a stochastic background, can be probed by the anisotropy pattern of the
CMB. The analysis of the gravitational effect on the CMB anistropies angular power spectra in temperature and polarization with recent Planck \citep{Akrami:2018vks}, BICEP/Keck Array \citep{Ade:2018iql}, SPT \citep{Keisler:2015hfa} data strongly constrains fields with amplitude values larger than a few co-moving $\sim  ~\rm nG$ on $\sim$ Mpc scales \citep{Paoletti:2019pdi,PLANCK2015,Zucca:2016iur,Paoletti:2012bb,Paoletti:2010rx,Shaw:2010ea}. 
Corresponding bounds have been also derived for an homogeneous primordial magnetic field, whose main additional effect is the breaking of isotropy, that is constrained to very few nG already with COBE satellite data \citep[][]{1997PhRvL..78.3610B}. However, these bounds may be slightly relaxed in the presence of free-streaming neutrinos \citep[][]{Adamek:2011pr}.
On the other hand,  magnetic fields in cosmic voids are constrained to be larger than the lower limits deduced by the absence of an Inverse Compton Cascade from distant blazars \citep[e.g.][]{2009ApJ...703.1078D,2010Sci...328...73N,2011ApJ...727L...4D,2014ApJ...796...18A,2015PhRvD..91l3514C,2015PhRvL.115u1103C}, of order $\sim 10^{-7} \ \rm nG$ {\footnote{See however \citet{2012ApJ...752...22B} for a possible different interpretation.}}.
Even more stringent limits on the amplitude of magnetic seed fields have been derived by including post-recombination 
heating \citep{Kunze:2014eka,Chluba:2015lpa,Paoletti:2018uic} or by 
modelling the small-scale baryonic density fluctuation induced by primordial magnetic fields,
leading to inhomogeneous recombination and heating, which would alter the peaks and heights of the large-scale anisotropies of the CMB \citep[][]{2018MNRAS.481.3401T,2019PhRvL.123b1301J}.

Any significant detection of magnetic fields beyond galaxies and galaxy clusters (or even any robust upper limit) would thus help to explore the origin of cosmic magnetism, because several theoretical 
works have shown that the radio signatures of drastically different magnetic field scenarios would leave very different imprints in Faraday Rotation and/or synchrotron emission from the magnetised cosmic web \citep[][]{donn09,va15radio,va17cqg}.  Due to the very weak radio signal expected outside of the overdensities typical of halos and cluster outskirts, different observational proxies have also been proposed, i.e. 
 by studying the propagation of Ultra-High-Energy Cosmic Rays  \citep[e.g.][]{2005JCAP...01..009D,2017arXiv171001353H}, by using the non-detection of Inverse Compton  emission from distant blazars.
 From the latter, lower limits of $\geq 10^{-16} ~ \rm G$ on $\sim \rm ~Mpc$ scales have been derived \citep[][]{2009ApJ...703.1078D,2010Sci...328...73N,2011ApJ...727L...4D,2014ApJ...796...18A,2015PhRvD..91l3514C,2015PhRvL.115u1103C}, and more stringent limits will be obtained in the near future by  the Cherenkov Telescope Array (CTA)
\citep[][]{2013IAUS..294..459S,2016ApJ...827..147M}.

Finally, the presence of significant large-scale magnetic fields has  been suggested as a possible explanation for the puzzling lack of infrared absorption in the observed spectra of distant blazars \citep[e.g.][]{2012PhRvD..86h5036T,2012PhRvD..86g5024H}.  
Axion-like particles (ALPs)  are promising candidate for DM \citep[][]{1988PhRvD..37.1237R,2003JCAP...05..005C} and they can oscillate into high-energy photons (and back) in the presence of background magnetic fields, reducing the effective opacity of emitted $\gamma$-ray photons \citep[][]{2012PhRvD..86g5024H}. With recent work we simulated the propagation of photons from redshift $z=1$ and computed the expected conversion into  ALPs, and we found that photons-ALPs oscillations are possible for lines of sight crossing structures with $\sim 1-10 ~\rm nG$ on scales of a few $\sim \rm Mpc$. \citep[][]{2017PhRvL.119j1101M}, again likely to be tested by the upcoming CTA \citep[][]{2017PhRvL.119j1101M}.\\

In this work, we present for the first time cosmological simulations of primordial magnetic fields derived from the constraints from the CMB. By evolving different primordial magnetic field models from $z=40$ to $z=0$, we generated observable signatures across the cosmic web and used them to investigate whether further constraints can be put on primordial scenarios.  This paper is structured as follows: after describing our simulations and numerical tools in Sec. 2, we present our results in Sec. 3. We provide physical and numerical caveats in Sec. 4 before we summarise and conclude our work in Sec. 5.

\begin{figure}
\includegraphics[width=0.45\textwidth]{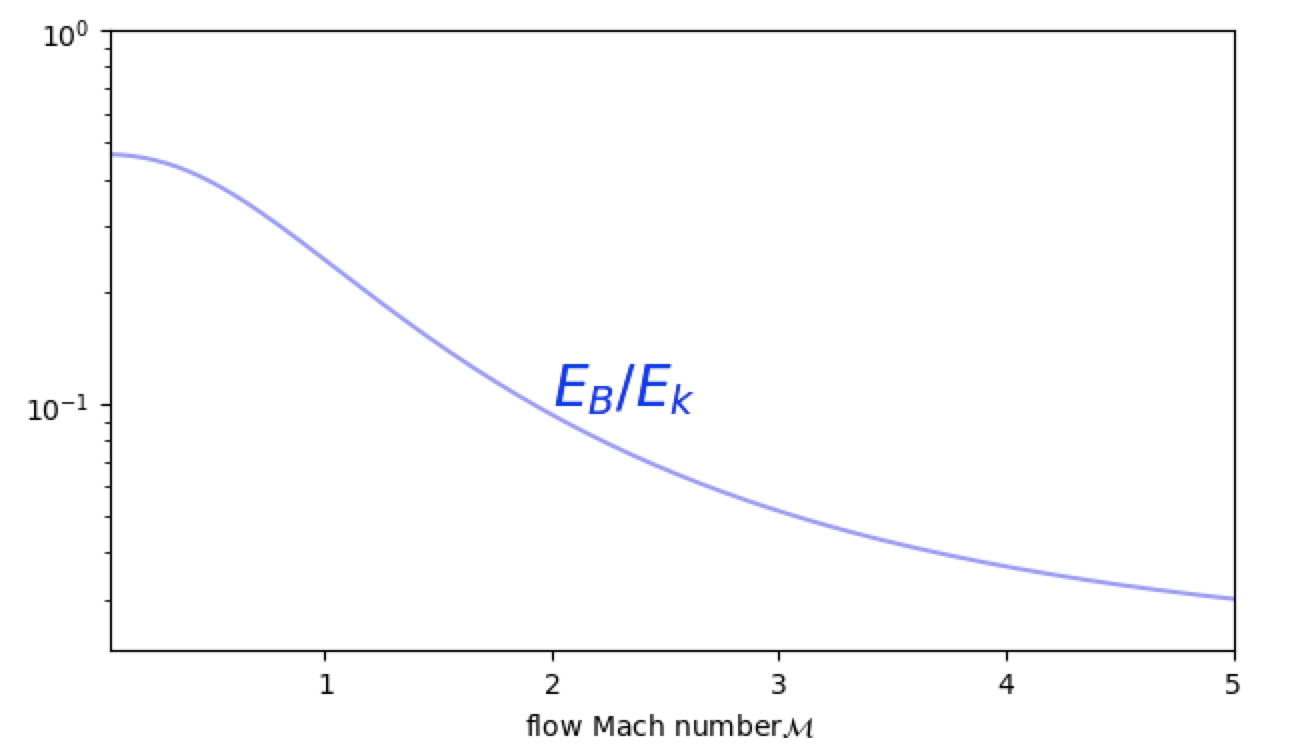}
\caption{Predicted saturation level of magnetic energy with respect to the turbulent kinetic energy, in the case of solenoidal forcing of turbulence, as computed in \citet{fed14} and adopted in our sub-grid modelling of small-scale dynamo amplification (Sec.2.1.2).}
\label{fig:federrath}
\end{figure}  

%
\begin{table*}
\label{tab_cosmo}
\caption{List of cosmological {\enzo} simulations produced for this work.}
\centering \tabcolsep 5pt 
\begin{tabular}{c|c|c|c|c|c|c|c|c|c|c|c}
Run ID & Volume  & Resolution & $\Omega_M$ &  $\Omega_b$  &  $\Omega_\Lambda$ & $\sigma_8$ & $h$ & $k_D$ &$B_{\rm Mpc}$ & $B_{\rm eff}$ &description\\
       &   [Mpc]    &   [kpc]     &       &      &     &     & [$100 ~\rm km/s/Mpc$] & [1/Mpc] & [nG] & [nG] & \\ \hline
B0 & 100  &  195 &  0.308 & 0.0478 & 0.692 & 0.815 & 0.678 &  -     & 2.0 & 2.0 &  homogeneous \\
B1 & 100  &  195 &  0.308 & 0.0478 & 0.692 & 0.815 & 0.678 & 84.7 & 2.0 & 2.53  &  $\alpha=-2.9$ \\
B2 & 100  &  195 &  0.308 & 0.0478 & 0.692 & 0.815 & 0.678 & 25.8 & 1.87 & 47.63 &  $\alpha=-1.0$ \\
B3 & 100  &  195 &  0.308 & 0.0478 & 0.692 & 0.815 & 0.678 & 37.6 & 0.35 & 70.1 & $\alpha=0.0$ \\
B4 & 100  &  195 &  0.308 & 0.0478 & 0.692 & 0.815 & 0.678 & 56.6 & 0.042 & 95.21 & $\alpha=1.0$ \\
B5 & 100  &  195 &  0.308 & 0.0478 & 0.692 & 0.815 & 0.678 & 87.9 & 0.003 & 119.21 & $\alpha=2.0$ \\
\end{tabular}
\end{table*}

\begin{figure}
\includegraphics[width=0.45\textwidth]{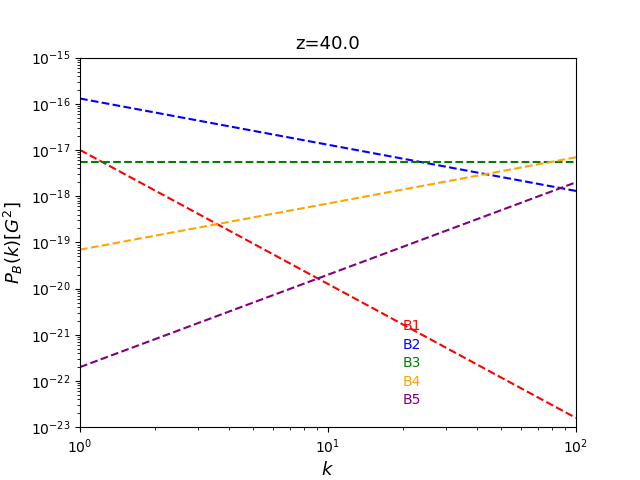}
\caption{Input 3-dimensional power spectra ($z=40$) of magnetic fields for our models.}
\label{fig:pk0}
\end{figure}

\begin{figure}
\includegraphics[width=0.45\textwidth]{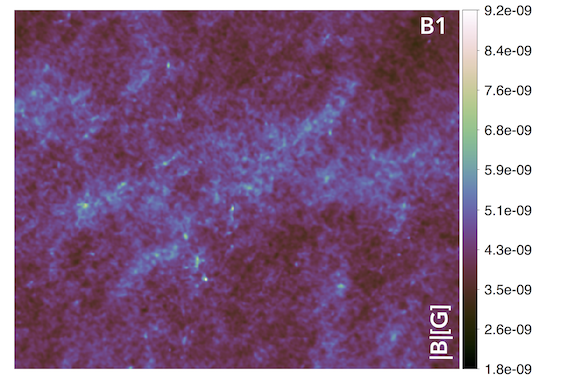}
\includegraphics[width=0.45\textwidth]{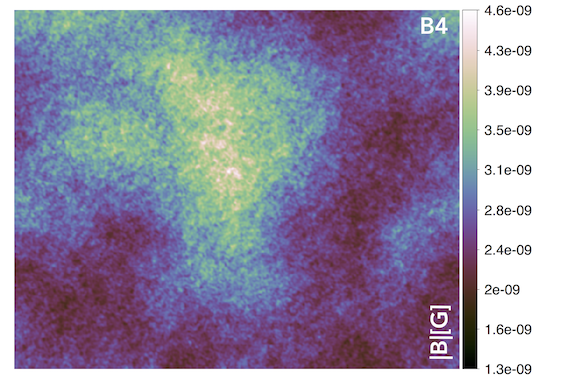}
\caption{Projected (gas mass-weighted) physical magnetic field  strength along the entire 100 Mpc line of sight for run B1 and B4 at the begin of our simulations ($z=40$)}.
\label{fig:mapB_z40}
\end{figure}

\begin{figure*}
\includegraphics[width=0.99\textwidth]{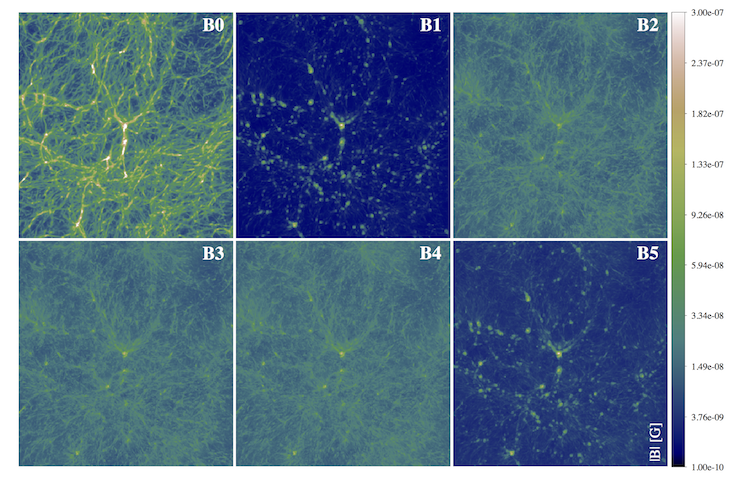}
\includegraphics[width=0.99\textwidth]{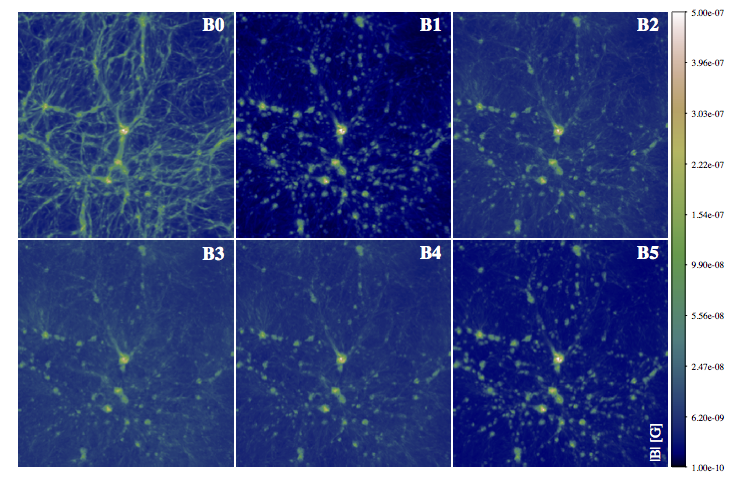}
\caption{Projected (gas mass-weighted) proper magnetic field  strength along the entire 100 Mpc line of sight, for all our models at $z=1$ (top) and at $z=0$ (bottom). }
\label{fig:mapB_z1}
\end{figure*}

\section{Methods}
\label{sec:methods}

\subsection{Cosmological Simulations}
\label{subsec:sim}

We used the cosmological Eulerian code  {\enzo} \citep{enzo14} to resimulate a cosmic volume of (100 Mpc)$^3$ comoving, with the constant spatial resolution of $\Delta x=195$ comoving kpc, using $512^3$ cells and Dark Matter particles.  We investigate the evolution and topology of magnetic fields in the most rarefied cosmic regions, removed from the contamination from galaxy-related processes. In order to make up for the lack of resolution, we apply a subgrid scheme for dynamo amplification (Sec.\ref{dynamo}). 

The adopted cosmology is a $\Lambda$CDM model with $\Omega_b=0.0468$, $\Omega_m=0.308$, $\Omega_{\Lambda}=0.692$, $H_0=67.8 \rm ~km/s/Mpc$, $\sigma_8=0.815$, and a spectral index for the initial matter power spectrum of $n=1.0$ \footnote{Due to the little correlation of the scalar spectral index with the primordial magnetic field configuration we expect little dependence on the value of ns of the field final properties.}
and an initial redshift $z_{\rm ini}=40$. 

\subsubsection{MHD method}
\label{mhd}
The adopted scheme for magneto-hydrodynamics (MHD) is the conservative Dedner formulation \citep[][]{ded02}, which uses hyperbolic divergence cleaning to keep the $\nabla \cdot \vec{B}$ as small as possible, in combination with the Piecewise Linear Method reconstruction (PLM) technique and with the 
 Harten-Lax-Van Leer (HLL) approximate Riemann solver. The time integration is based on the total variation diminishing (TVD) second-order Runge-Kutta (RK) scheme \citep[][]{1988JCoPh..77..439S}.  
 
 The choice of using a constant spatial resolution, rather than an adaptive one, is related to the fact that 
 all effects we are looking for in this project are related to underdense cosmic regions (e.g. voids) or mildly dense structures (e.g. matter sheets, filaments, cluster outskirts), and therefore the use of a fixed spatial resolution is helpful for our analysis. For applications of adaptive mesh refinement for resolving turbulence in the intergalactic medium, we refer the reader to  \citet{2017MNRAS.469.3641I}.
 Tests on the dependence of some of our results on the spatial resolution are presented in the Appendix. 
 
 \subsubsection{Subgrid Dynamo Model}
 \label{dynamo}
 
 Our simulations are non-radiative, i.e. only the effects of cosmological expansion, gravity and magneto-hydrodynamics are included. To make up for the lack of small-scale dynamo amplification in the turbulent interiors of halos, we applied a subgrid model for magnetic field amplification \citep{po15,bm16,wi17} within overdense regions ($\rho \geq 50 \langle \rho \rangle $)\citep[e.g.][]{va17cqg}. At run time we measure the gas vorticity and use it to estimate the dissipation rate of solenoidal turbulence into magnetic field amplification. Our procedure is based on \citet{fed14}, and on their fitting formulas to predict the growth of magnetic energy from scales that our simulation cannot directly resolve. \\
 
 The local velocity enstrophy, $(\nabla \times \vec{v})^2 \equiv \epsilon_{\omega}$ is a convenient quantity to estimate the turbulent dissipation rate, because the turbulent kinetic energy flux is conserved along the cascade. Following previous work \citep[][]{jones11,po15,va17turb,wi17b}, we assume that a small fraction, $\eta_t \approx 10^{-2}$, of such kinetic power gets channeled into the amplification of magnetic fields,  $F_{\rm turb} \simeq \eta_t \rho \epsilon_{\omega}^3/L$, where $L$ is the stencil of cells to compute the vorticity.  The fraction of turbulent kinetic power that gets converted into magnetic energy, $\epsilon_{\rm dyn}$, sets the amplified magnetic energy as $E_{\rm B,dyn} = \epsilon_{\rm dyn}(\mathcal{M})F_{\rm turb} \Delta t$. 
 For a reasonable guess on $\epsilon_{\rm dyn}$, we  
 rely on \citet{fed14}, who simulated small-scale dynamo in a variety of  conditions for the forcing of turbulence. We can thus estimate the saturation level and the typical growth time of magnetic fields as a function of the local Mach number of the flow ($\mathcal{M}$), and set $\epsilon_{\rm dyn}(\mathcal{M}) \approx (E_B/E_k) \Gamma \Delta t$, where $E_B/E_k$ is the ratio between magnetic and kinetic energy at saturation, and $\Gamma$ is the growth rate, taken from \citet{fed14}. 
In Fig.~\ref{fig:federrath} we show the $E_B/E_k (\mathcal{M})$ relation adopted in our model, which is derived for a purely solenoidal forcing of turbulence  \citep{fed14}, which is appropriate for halos, given the predominance of solenoidal motions \citep[e.g.][]{miniati14,va17turb}. As a reference value, for a $\mathcal{M}=0.5$ turbulent Mach number, the model predicts saturated level of magnetic energy which is $\approx 40 \%$ of the local solenoidal turbulent kinetic energy.
 In this approach, we need to specify the topology of the amplified magnetic field. For simplicity, the additional field is taken to be parallel to the local gas vorticity, such that the newly generated field is solenodial by construction. Energy and momentum are conserved assuming an isotropic dissipation of the small-scale velocity vectors. Admittedly, this procedure is simpler than more sophisticated subgrid models \citep{gr16}. However, this simplistic method reproduced the results obtained by other methods \citep[e.g.][]{ry08,va17cqg,hack19}.
 While the application of the sub-grid dynamo model adds realism to the magnetic field distribution in our volume, it has no impact on the intermediate and low-density regime in which we study the differences introduced by our CMB-based magnetic fields. As it solely relies on the gas velocity field, which is basically identical in all runs, the contribution from the sub-grid dynamo model within halos is exactly the same in all models. 
 
As an important caveat to our sub-grid dynamo model, we stress that in reality the dynamo growth rates must depend on the magnetic Prandtl number ($P_M$),  on the kinematic Reynolds number ($R_e$), as well as on the turbulence driving \citep[][]{2011PhRvL.107k4504F,fed14}. For example, \citet{2012PhRvE..85b6303S} measured the increase of the growth rate with the (numerical) Reynolds number. Furthermore, the Reynolds and the Prandtl numbers in the ICM are expected to be  much larger than anything that can be resolved by MHD simulations in the near future \citep[e.g.][]{bl07,review_dynamo}. As usual in cosmological MHD simulations, $P_M \approx 1$, i.e. the kinematic and the magnetic Reynolds number are equal as we only have numerical viscosity and resistivity   \citep[e.g.][]{review_dynamo}. 
 Of course, given the complexity of plasma physics, this convenient assumption is highly questionable  \citep[e.g.][]{2004ApJ...612..276S,bl11b,2016ApJ...817..127B}. 
   Moreover, our model misses turbulence in the early Universe that develops during the collapse of filaments and which is generated on scales too small to be resolved. In this case, the main driving mode of turbulence should be compressive driving \citep[e.g.][]{2018PhT....71f..38F}.
   In this respect, more flexible sub-grid model of dynamo amplification may be explored in the future \citep[e.g.][]{2017PhRvE..95c3206G,2019MNRAS.487.4525G}, even if the main conclusions of our paper, regarding observable properties of cosmic structures on $\geq \rm Mpc$ scale, are expected to remain unaltered.

 \subsubsection{Primordial magnetic field models}
We modelled the primordial magnetic fields using a stochastic background, described by the two-point correlation function \citep{Mack:2001gc,Finelli:2008xh}:
\begin{equation}
\langle B^\star_i({\mathbf k})B_j({\mathbf k'})\rangle =
\delta^{(3)} ({\mathbf k}-{\mathbf k'}) P_{ij}({\mathbf{\hat k}})
P_B(k) (2\pi)^3,
\end{equation}
where $i$ and $j$ are spatial indices. $\delta^{(3)}({\mathbf k}-{\mathbf k'})$ is the Dirac delta function, with unit vector 
$\hat{k}_i=k_i/k$, $P_{ij}({\mathbf{\hat
k}})=\delta_{ij}-\hat{k}_i\hat{k}_j$ is the operator for transverse plane projection, and $P_B(k)$ is the power spectrum of the magnetic field.
The fields scale dependence is described with a power law spectrum: $P_B(k) = P_{B0}k^{\alpha} $ characterised by the amplitude of the fields and the spectral index. As a convention, we describe the amplitude by smoothing the fields on the scale $\lambda$ and using $B_\lambda$:
\begin{equation}
P_B(k) = P_{B}k^{\alpha}= \frac{2\pi^2 \lambda^3
B^2_\lambda}{\Gamma(n_B/2+3/2)} (\lambda k)^{\alpha},
\label{energy-spectrum-H}
\end{equation}
where $\lambda$ is a  comoving smoothing length (with spatial frequency $k_\lambda=2\pi/\lambda$), with a Gaussian kernel $\propto \mbox{exp}[-x^2/\lambda^2]$.
For $k>k_D$, where $k_D$ is the cutoff wavenumber, the power spectrum gets dissipated through  Alfv\'en wave damping. 
The values of $k_D$ for each magnetic field model is given in Tab.1 (8th column).

It is customary to set $\lambda=1 \rm ~ Mpc$, and therefore in the remainder of the paper we will use 
$B_{\rm Mpc}=B_\lambda$ to refer to the smoothed magnetic field amplitude. 

In Table 1, we provide both the magnetic field settings and  the cosmological parameters used in the simulations. 
The primordial magnetic fields we assume are a bunch of different cases from the range of possible spectral indices, from the almost scale invariant $\alpha=-2.9$ whose energy momentum tensor in Fourier space is infrared dominated, to an high values of $\alpha=2$ which corresponds to the minimum index allowed for causally generated magnetic fields.  
The values for the amplitudes of the fields are derived by the constraints with the combination of the most recent data of Planck 2018 \citep{Akrami:2018vks} with ground based observatory as BICEP/KECK \citep{Ade:2018iql} and the South Pole Telescope \citep{Keisler:2015hfa} following \cite{Paoletti:2019pdi} using the gravitational scalar, vector and tensor effects of PMFs on CMB anisotropies.\footnote{The additional constraints that are not derived in  \cite{Paoletti:2019pdi} have been computed with the same approach and code as \cite{Paoletti:2019pdi}}. We stress that, in this work, the magnetic seeds used are modelled with configurations, amplitudes and corresponding scale dependences, that are directly derived from CMB data constraints. 
Concerning the homogeneous field case, we assume a value of 2 nG in agreement with the COBE constraints \citep[][]{1997PhRvL..78.3610B}, that still remain the tightest constraints on homogeneous field from CMB anisotropies. This value is pretty generous and stronger constraints should be provided by Planck in the future, but are not currently available therefore we refer to the COBE ones.
Note that the cosmological parameters are not varied for the different settings of the magnetic fields because, as demonstrated by the constraints derived with different observational data and the forecasts for future experiments  \citep{Paoletti:2010rx,Paoletti:2012bb,PLANCK2015,Paoletti:2019pdi}, the inclusion of primordial magnetic fields does not introduce strong degeneracies with the standard cosmological parameters.

We can define an "effective" magnetic field, corresponding to the magnetic energy density generated by the fields above, $B_{\rm eff}=\sqrt{\langle B^2 \rangle}$ where the latter is the integrated square magnetic field up to $k_D$ as in \citet{Finelli:2008xh} and \citet{2013ApJ...770...47K}, 
it can be computed as: 

\begin{equation}
B_{\rm eff}=B_{\rm Mpc} \cdot \frac{(k_D \lambda)^{(\alpha+3)/2}}{\sqrt{\Gamma(\alpha/2+5/2})}.    
\end{equation}

The $B_{\rm eff}$ corresponding to each model is given in the 10th column of Tab.1:  while this is the effective magnetic field value expected if all scales down to $l_D \approx 1/k_D$ can be probed, our simulations can only resolve a larger spatial resolution coarser than $l_D$. 

As a consequence, the effective magnetic field amplitude that can be captured by our simulations will be typically smaller that $B_{\rm eff}$. 
In this work we only consider non-helical magnetic fields. Helical magnetic fields can be generated during inflation \cite{Field:1998hi,Vachaspati:2001nb,Sigl:2002kt,Sharma:2018kgs} and helicity represent an important component in the evolution of the fields as shown by simulations \citep{Christensson:2000sp,Christensson:2002xu,Saveliev:2013uva,Kahniashvili:2016bkp,Brandenburg:2016odr}. The helicity impact on CMB anisotropies has been investigated in \citep{Pogosian:2002dq,Caprini:2003vc,Kunze:2011bp,Ballardini:2014jta,Kahniashvili:2014dfa}, and the constraints on the helical field amplitude have beebn explored in \citep{PLANCK2015}. Due to the complexity degree required by the presence of an helical component in the magnetic fields, we leave the helicity treatment for a future work. 

 Here we impose the initial conditions for the magnetic fields described above at the start of runs, $z_{\rm ini}= 40$. The elapsed time between recombination and $z = 40$ the two epochs is $\approx 65 \rm ~Myr$, which is negligible compared to the total time of the simulation, and also no significant structure formation is expected to take place in between these two epochs, hence this simplification does not pose any problem for our analysis\footnote{{For the current work we do not consider the additional post recombination effect of PMFs.}}. 

We note that in our analysis we consider only the PMFs as a stochastic background with a given a spectral index and an amplitude compatible with the CMB anisotropies constraints, leaving  the additional impact of magnetized cosmological perturbations on the matter power spectrum  (see for example \citealt{Sethi:2004pe,Finelli:2008xh,Shaw:2009nf,Fedeli:2012rr,2013ApJ...770...47K}) for a future work.

Fig.~\ref{fig:pk0} gives the input magnetic power spectra for our models, while Fig.~\ref{fig:mapB_z40} gives the example of the projected maps of magnetic fields strength at $z=40$ for the B1 ($\alpha=-2.9$) and for the B4 ($\alpha=1.0$) models. We remark that the variation of the amplitude with the configuration is central for the results we will comment in the next Sections:  spectra which increase the power on the smallest scales are already strongly constrained by the low amplitude of the CMB, whereas fields with more infrared spectra are allowed to larger amplitudes but lower  power on much larger scales. The interplay between the scale dependence of the fields, and the constraints on their amplitude by the CMB, will be reflected into the lower redshift constraints explored in the rest of this work.

\begin{figure}
\includegraphics[width=0.4\textwidth]{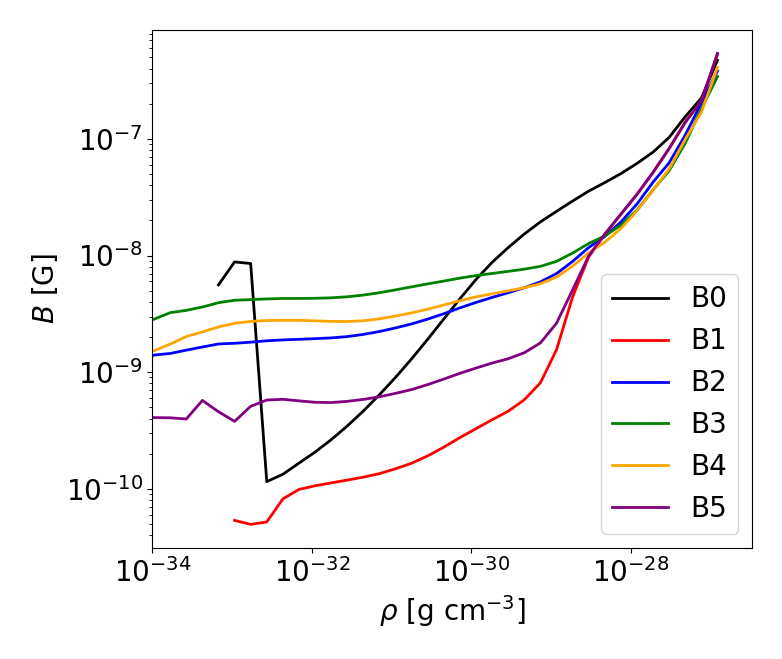}
\includegraphics[width=0.4\textwidth]{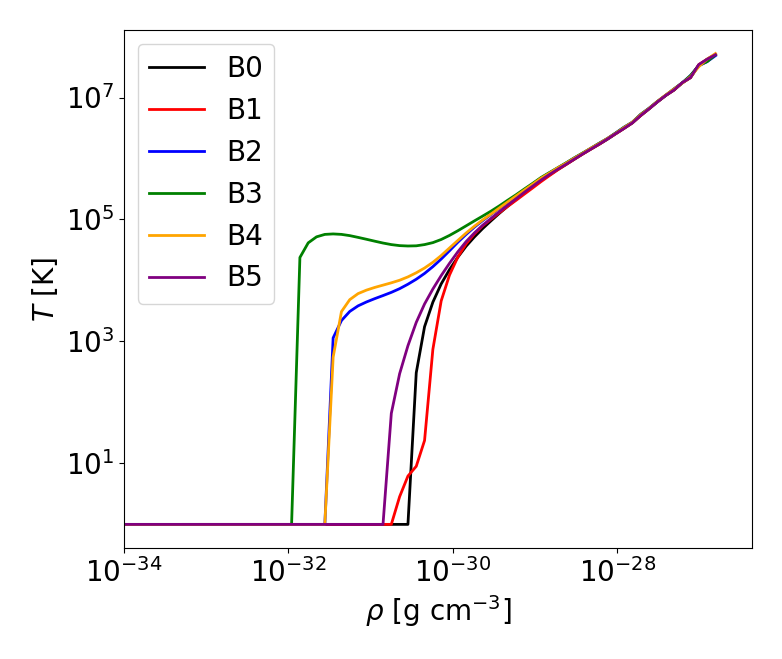}
\caption{Median magnetic field strength as a function of baryon density (top) and median gas temperature for the same gas density bins (bottom) for all models at $z=0.02$. }
\label{fig:rhoB}
\end{figure}

\begin{figure*}

\includegraphics[width=0.9\textwidth]{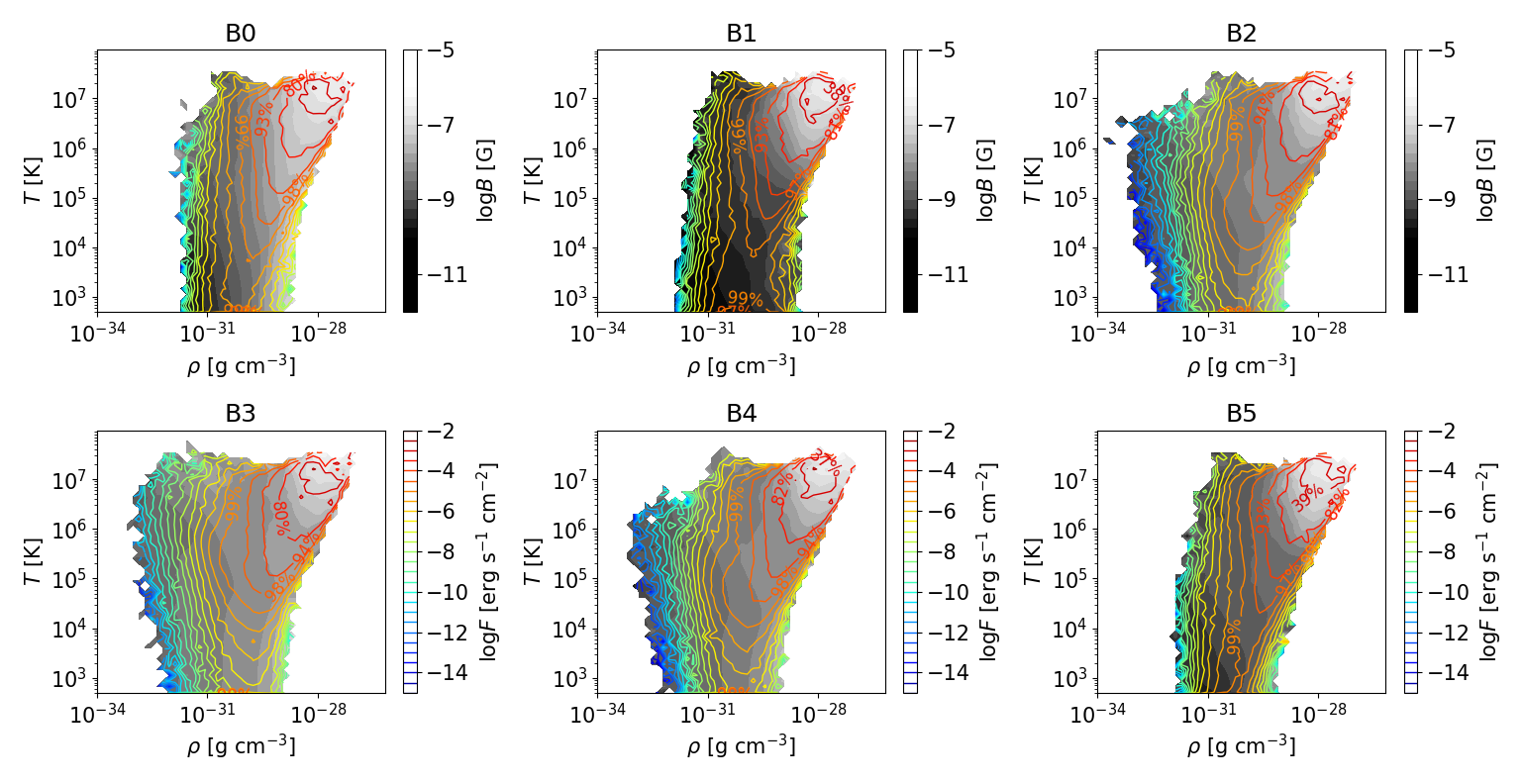}
\caption{Phase diagrams of baryons in all simulations at $z=0.02$. The additional color coding gives the median magnetic field strength in each gas phase (BW color palette) and the total dissipated energy flux at shocks (contours). }
\label{fig:phase}
\end{figure*}

\begin{figure}
\includegraphics[width=0.4\textwidth]{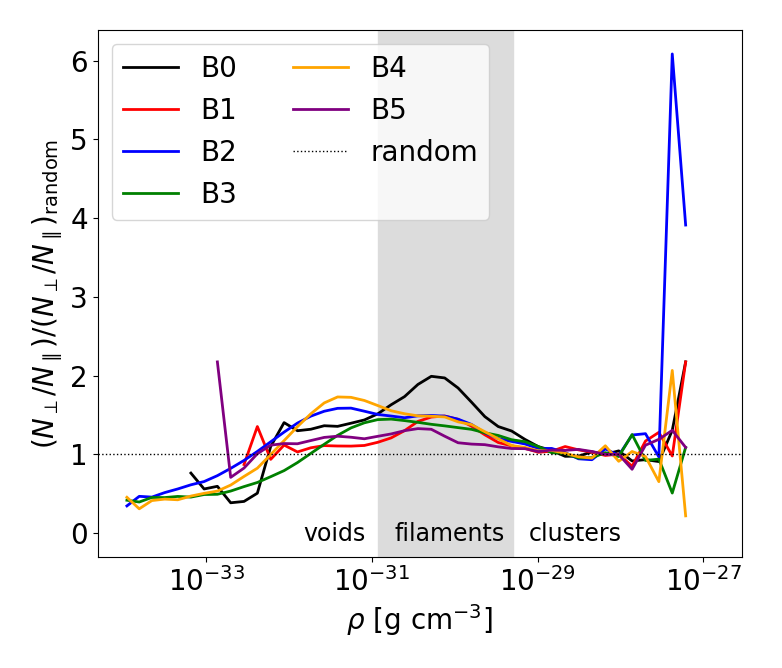}
\caption{Distribution of the ratio quasi-perpendicular to quasi-parallel shocks, normalised to the expected ratio in a random 3-dimensional distribution,  a function of gas density at $z=0.02$.}
\label{fig:obliq}
\end{figure}

\begin{figure*}
\includegraphics[width=0.33\textwidth]{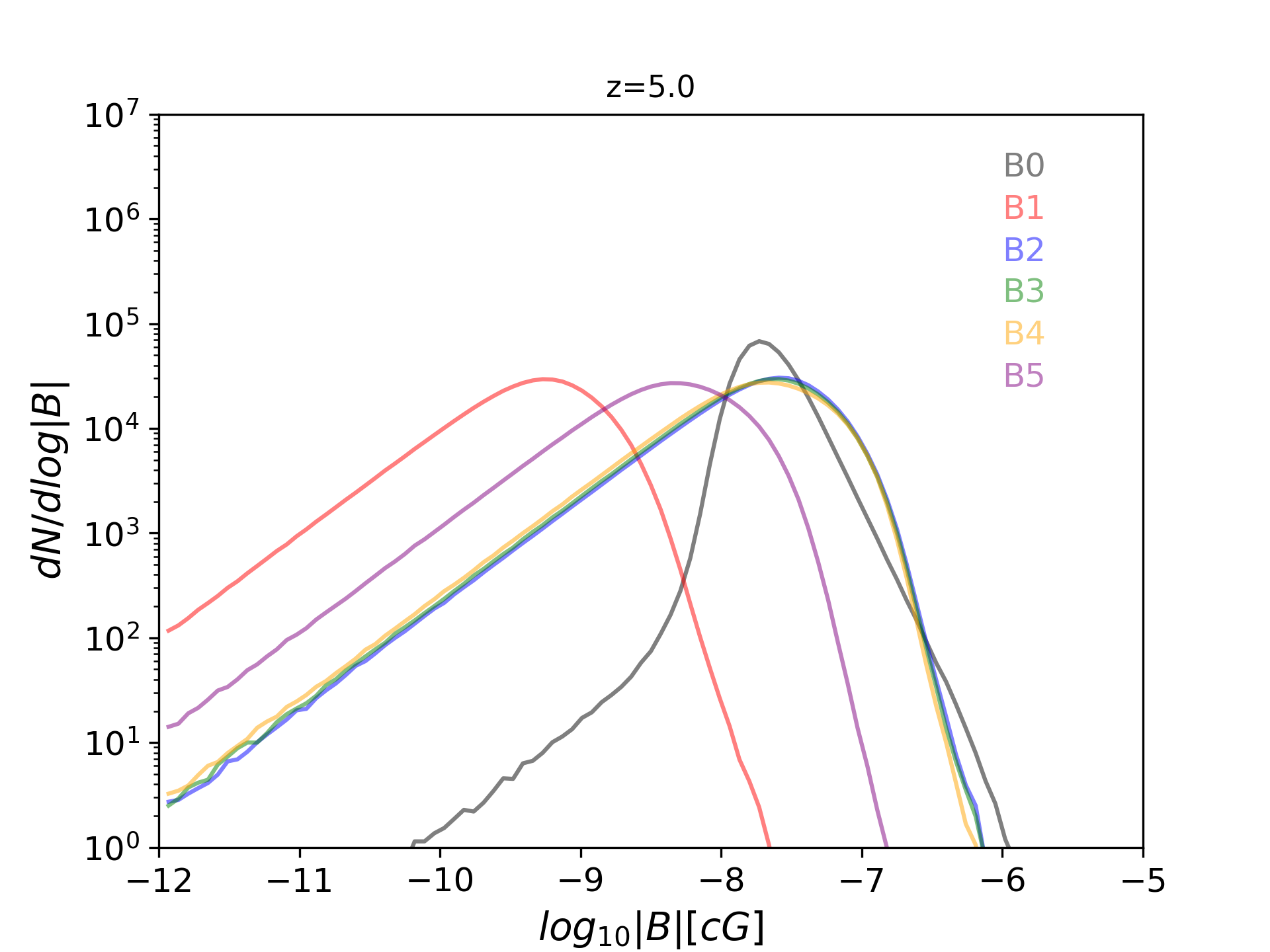}
\includegraphics[width=0.33\textwidth]{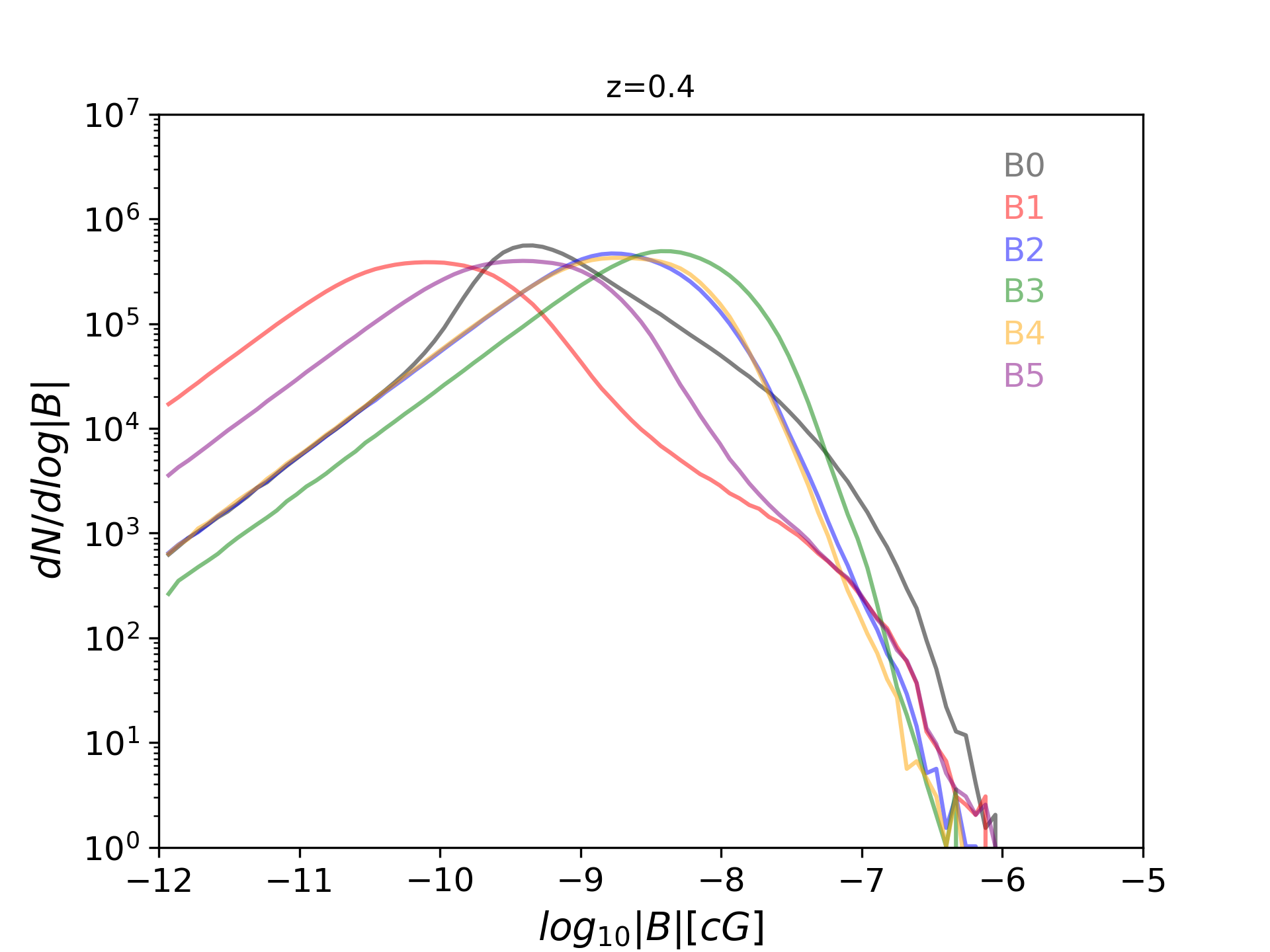}
\includegraphics[width=0.33\textwidth]{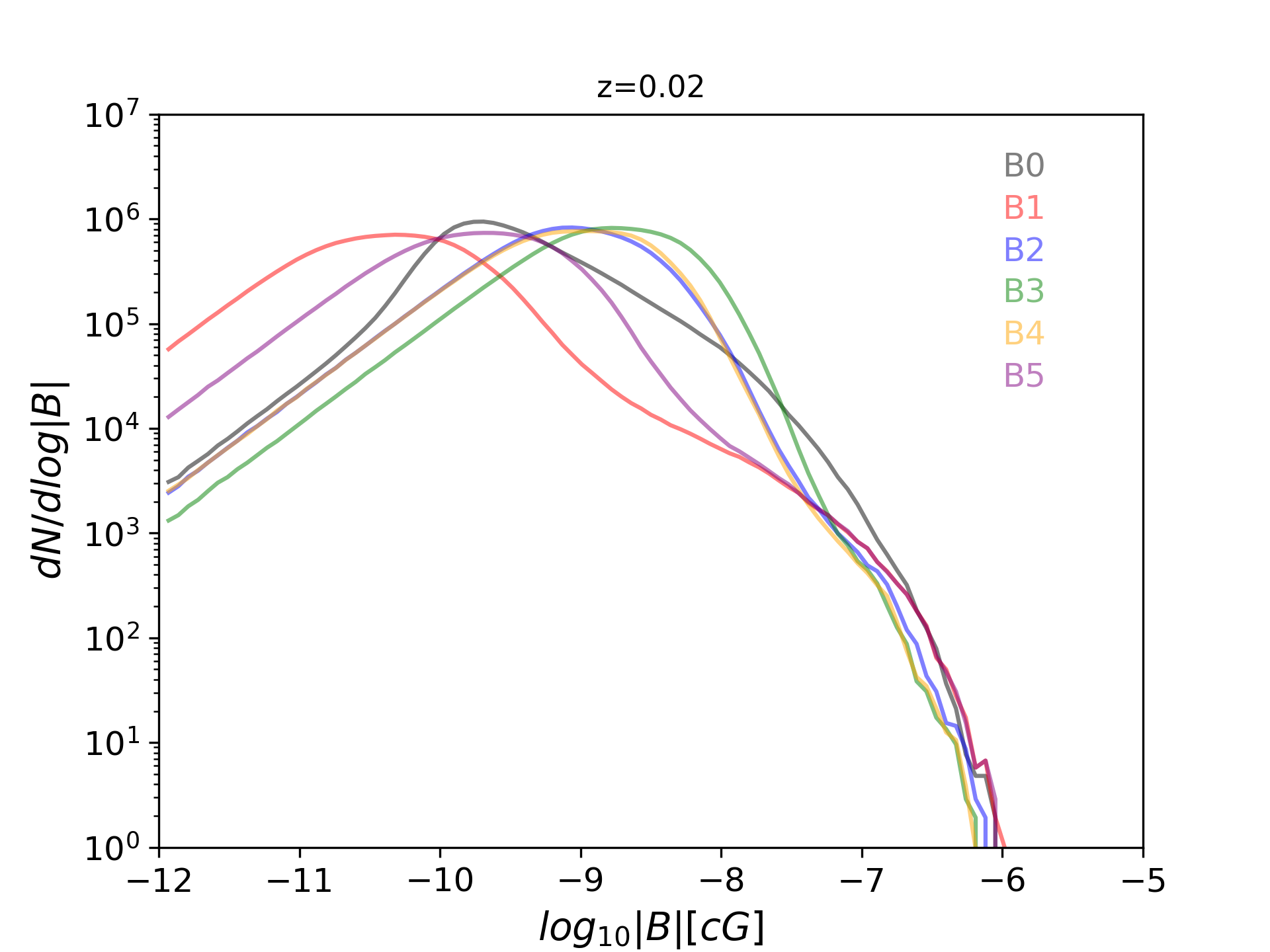}
\caption{Probability distribution functions of comoving magnetic field strength  for our models at  four different epochs.}
\label{fig:pdf}
\end{figure*}  

\begin{figure*}
\includegraphics[width=0.33\textwidth]{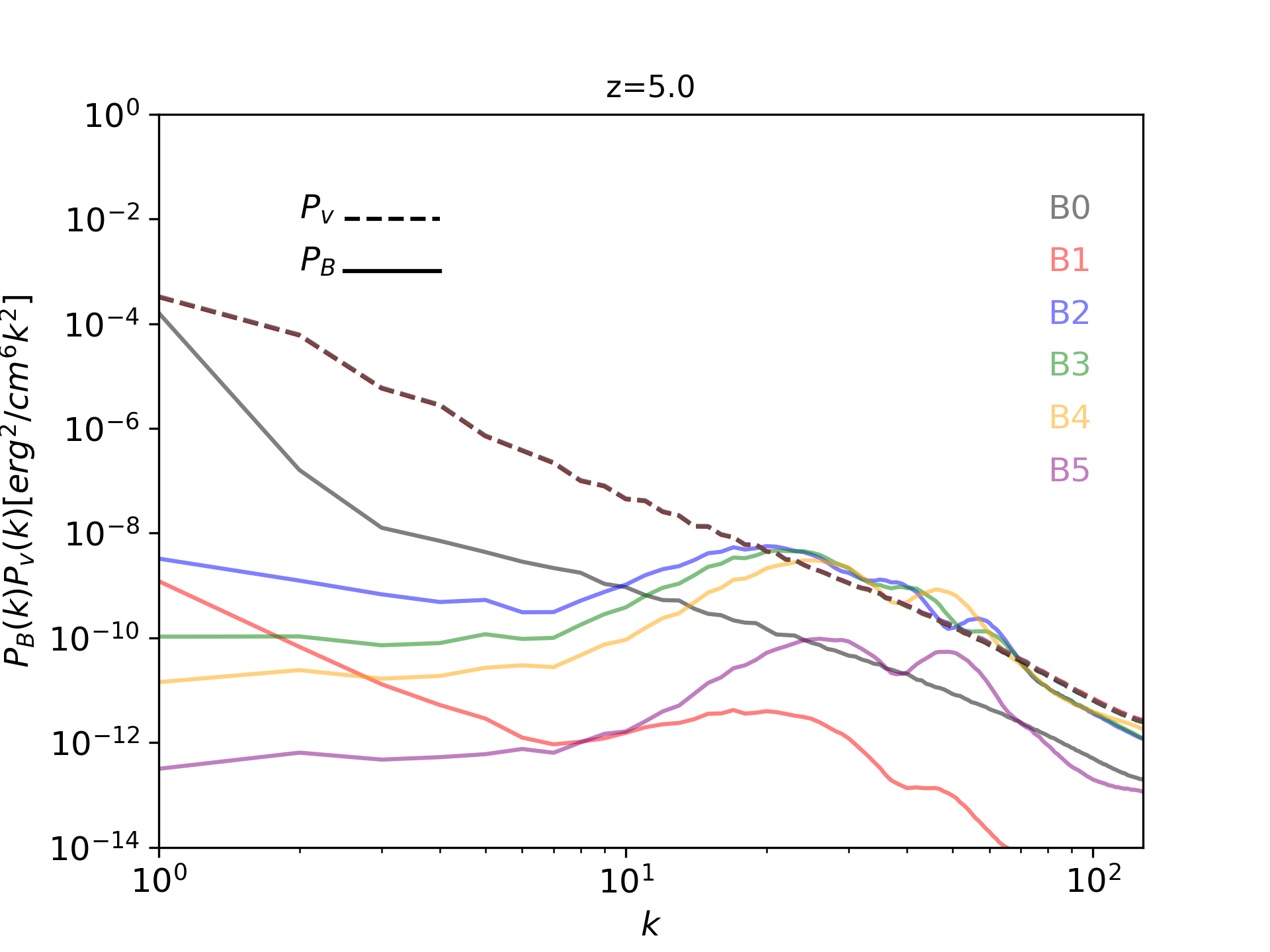}
\includegraphics[width=0.33\textwidth]{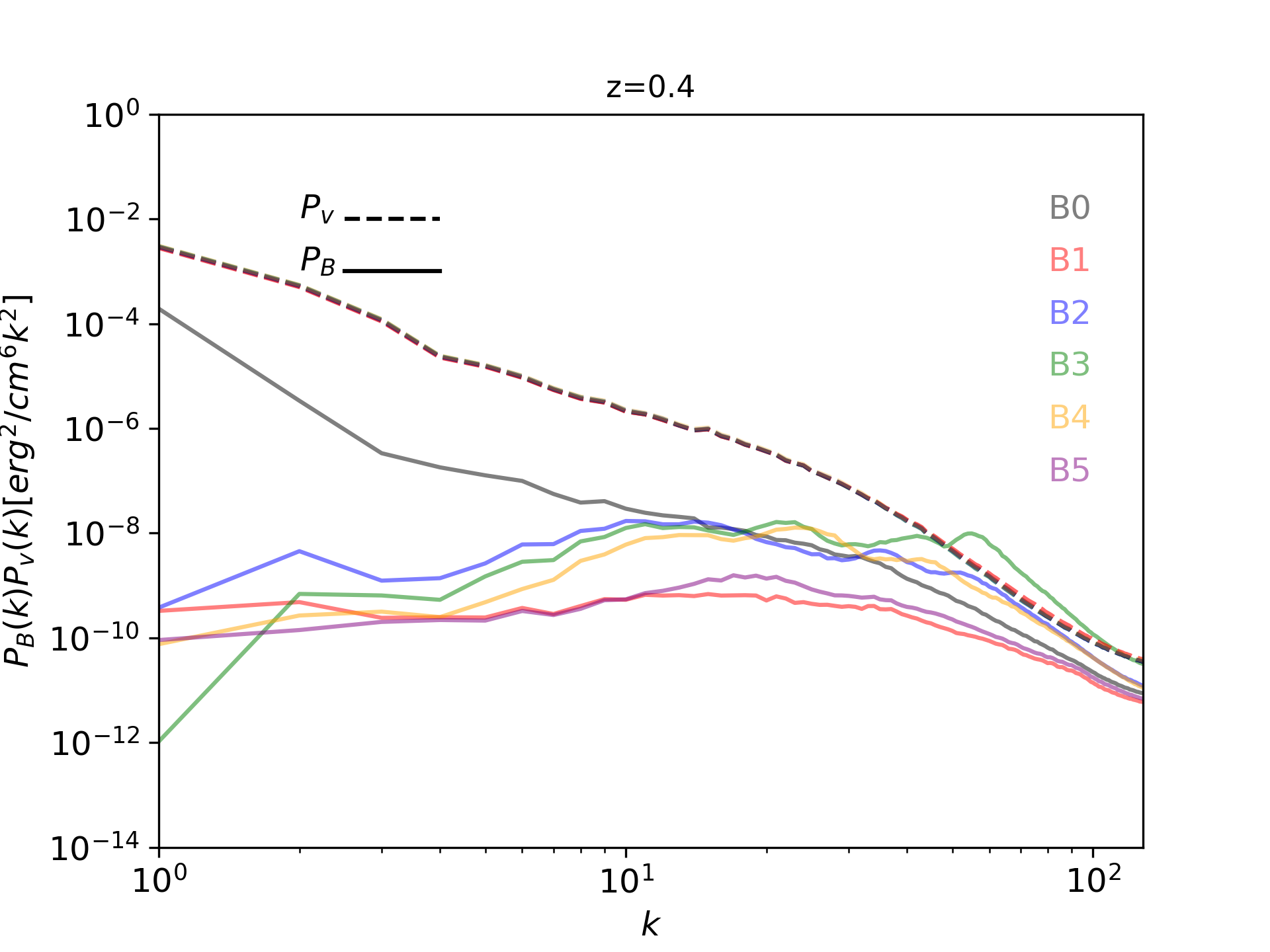}
\includegraphics[width=0.33\textwidth]{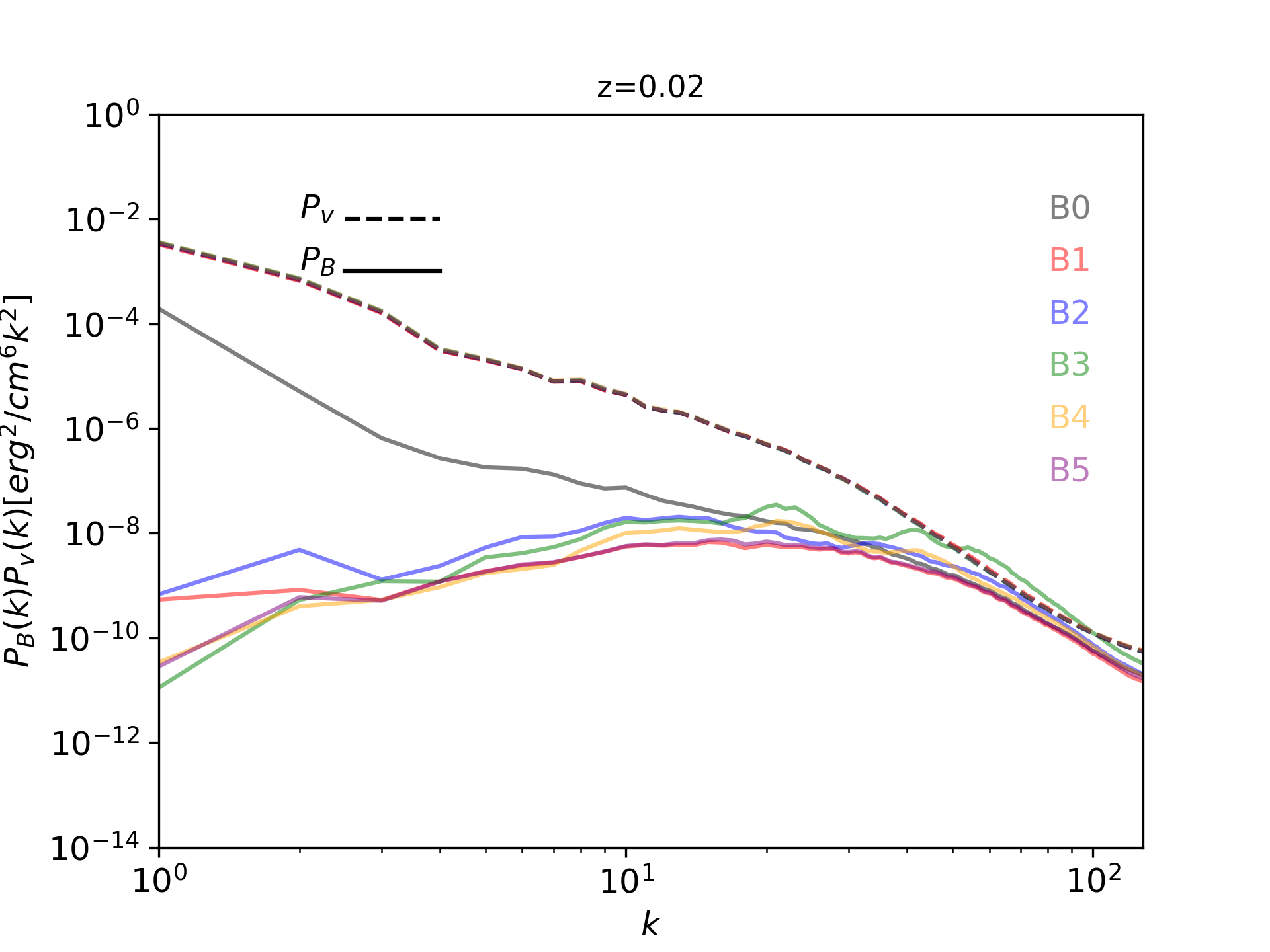}
\caption{3-dimensional magnetic power spectra (solid lines) and kinetic energy spectra (dashed lines) for our models at four different epochs.}
\label{fig:pk}
\end{figure*}

\begin{figure*}
\includegraphics[width=0.99\textwidth]{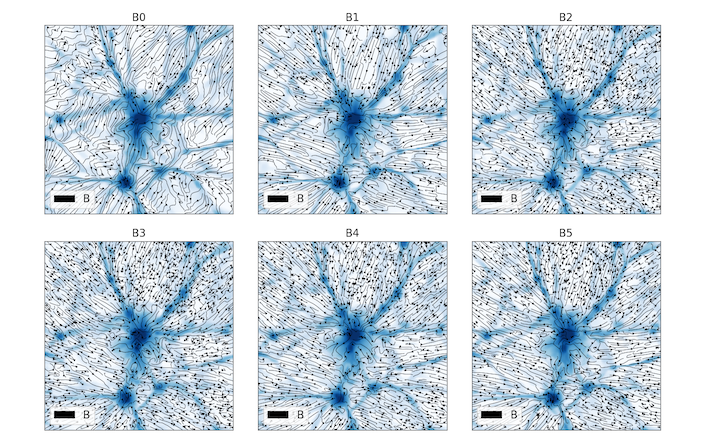}
\caption{Thin slices ($195$ kpc thick) showing the gas density (blue) and the magnetic field lines (black) for a $20 \times 20$ $\rm Mpc^2$ zoom around the most massive cluster in our simulated volume at $z=0$. }
\label{fig:zoom}
\end{figure*}  

\section{Results}
\label{sec:res}

\subsection{Properties of simulated magnetic fields}

\subsubsection{Global properties}

The panels in Fig.~\ref{fig:mapB_z1} show the spatial distribution of our simulated magnetic fields, for all models, at  $z=1$ and $z=0.02$.

As expected, large differences are present at all epochs, outside of collapsed matter halos in the volume. The amplitude of magnetic fields in the mildly dense (e.g. sheets/filaments) and underdense (e.g. voids) cosmic environment are found to vary by $\sim 10-10^2$ depending on the seeding model, event by the end of the simulation $z=0.02$. The magnetic field amplitude within halos is instead almost exactly the same in all cases, owing to our subgrid dynamo model, whose key ingredient is the small-scale gas kinetic energy, which is basically identical in all runs. 

In Fig.~\ref{fig:rhoB} we quantify the typical magnetic field level (and temperature) as a function of gas density. 
We can notice significant differences among models at low densities,  $\rho \leq 10^{-29} \rm g/cm^3$, i.e in the environment  of the outer regions of halos, or even outside them. 
In voids, there is a  $\geq 10$ difference in magnetic field intensity between models. 
While in the simple uniform seeding model (B0), the  magnetisation is a strong function of gas density, in most of the other models the presence of initial large fluctuations in the seed fields produces a broader distribution of magnetic fields in voids, smoothing out any strong dependence with density. 

A new finding of our analysis is the presence of differences in the average {\it gas temperature} in voids ($\rho \sim  10^{-32}-10^{-31} \rm g/cm^3$), depending on the magnetic seed model. For example, in models B2, B3 and B4 the average temperature of voids can reach $\sim 10^2-10^3 ~\rm K$, i.e. much beyond the $\leq 10 \rm ~ K$ of the standard B0 run here.  Both temperatures are unrealistic, in the sense that the reheating from reionization (which would rise the gas temperature to $\sim 10^3-10^4 \rm ~K$ everywhere, \citealt{hm96}) is completely missing in these models.  However, such differences are significant and hint at a substantial extra heating that primordial magnetic field fluctuations are introducing. 
Large initial fluctuations of magnetic field can become relevant for the local gas dynamics, because via induction equation they can trigger the formation of shock waves and the dissipation of turbulent motions into heat, in way qualitatively similar to what has been proposed during recombination epochs \citep[e.g.][]{2018MNRAS.481.3401T,2019PhRvL.123b1301J}. Also post-recombination, dissipative effects on the magnetic fields as MHD decaying turbulence and ambipolar diffusion, whose study is currently based mainly on approximated linear numerical treatments, show similar levels of the heating of the plasma  \citep{Kunze:2014eka,Chluba:2015lpa,Paoletti:2018uic}.
The development of extra shocks for gas with $\rho \leq 10^{-31} \rm g/cm^3$ is well captured by the $\rm (\rho,T)$ phase diagrams of Fig.~\ref{fig:phase}, in which we also computed the dissipated energy flux through shocks identified in the simulation {\footnote{Shocks are identified in post-processing, with a velocity-based scheme to compute the Mach number based on jumps of thermodynamical quantities, as in \citet{va09shocks} and \citet{Banfi20}}}. 
The additional shocks in  empty regions stand out in the  statistics of detected shocks in the left part of the phase diagrams, well away from the typical regime of  structure formation shocks driven by gravity \citep[e.g.][]{va11comparison}. 

We remark that such new classes of shocks driven by magnetic fields may not be physically very relevant, because of the aforementioned lack of a reionization heating floor in our model, as well as because the energy flux associated with such shock is extremely small, i.e. $\leq 10^{-7}$ of the total energy dissipation of kinetic energy in the total cosmic volume.\\

However, the exact topology of magnetic fields in the volume swept by cosmic shocks is also relevant as it can affect the acceleration efficiency of cosmic rays (CR). Protons and/or electrons should undergo different kinds of shock acceleration as a function of plasma parameters as well as of the topology of up-stream magnetic field \citep[e.g.][and references therein for a recent review]{Bykov19}. 
While CR protons should be efficiently accelerated by strong shocks with a quasi-parallel geometry between the shock normal and the upstream magnetic field via Diffusive Shock Acceleration (DSA), CR electrons may be accelerated in a two-phase fashion, in which they first gain energy via shock-drift acceleration if shocks are quasi-perpendicular, and are later suitable for acceleration by DSA \citep[e.g.][]{2014ApJ...783...91C,guo14}. 

Following \citet{Banfi20}, we measured  the shock obliquity from the shock propagation direction and the up-stream magnetic field $\mathbfit{B}$:
    $\theta=\arccos{\left(\frac{M_x\cdot B_x+M_y\cdot B_y+M_z\cdot B_z}{M\ B}\right)}$
which ranges from $0^{\circ}$ to $180^{\circ}$. For the identification of up-stream magnetic field values and of the local shock propagation direction, we rely on our velocity-based shock finder \citep[][]{Banfi20}.

Fig.~\ref{fig:obliq} shows the ratio between quasi-perpendicular ($|\theta| \geq 45^\circ$) and quasi-parallel ($\theta < 45^\circ$) for all models as a function of the gas density, normalized to the ratio that is expected from a purely random distribution of vectors in 3D space (which follows a simple $\propto \sin{\theta}$ probability distribution). 

All models present an excess of quasi-perpendicular shocks, with respect to a random distribution, in the $\rho \approx 10^{-32}-10^{-29}\ \mathrm{g\ cm^{-3}}$ density range. The excess is maximal at the overdensity typical of filaments, which is explained by the fact that filaments tend to stretch the local magnetic field perpendicular to the velocity shear and to the local density gradient \citep[][]{2017A&A...607A...2S}, while shocks mostly run along the density gradient \citep[][]{Banfi20}.
Therefore, in all models cosmic filaments are found in an environment in which the acceleration of CRs mostly proceed via quasi-perpendicular shocks, which would correspond to a reduced injection of CR protons in such structures \citep[see][for recent detailed analysis]{wittor20,Banfi20}. 

However, quantitative differences in the excess of quasi-perpendicular shocks can be seen, with the highest excess being present in the uniform B0 model. This can be understood in terms of magnetic {\it tension}, i.e. of the difficulty by the gas flow in bending magnetic field lines. The tension increases with the field curvature, hence it is not surprising that in the B5 model, where the tangling of magnetic field lines is maximal at small scales since the beginning, the gas dynamics struggle more than in the B0 case to bend the field direction and to align it perpendicular to the gas density gradient. Intermediate trends are found in the other models. As an effect of small-scale dynamo in halos, such effects are erased for $\rho \geq 10^{-28} \rm g/cm^3$, and the distribution of shock angles become purely random \citep[e.g.][]{wi17}.\\

Next, we measure the global volumetric and topological properties of magnetic fields for different epochs through their 
 Probability Distribution Functions (PDF) and their 3-dimensional power spectra, as shown by
Fig.~\ref{fig:pdf}- \ref{fig:pk},  where we considered three evolutionary steps as an example: $z=5,0.4$ and $0.02$. 

In the PDF, large  differences are seen at all epochs, with the tendency to similarity of all models in the high magnetisation regime ($\geq 0.1$ $\rm \mu G$) and for $z \leq 1$, after the small-scale dynamo has started operating in most halos. However, for the bulk of the volume of the cosmic web the relative differences in the peak of the PDF are preserved from $z=5.0$ to $z=0.02$, and they mirror the differences at $z=40$, suggesting that in all cases the overall evolution of the PDFs is mostly driven by compression-rarefaction. This further supports that, at least in theory, it should be possible to constrain the magnetisation model of our Universe, provided that we can probe the volumetric distribution of extragalactic magnetic fields today.

We computed the power spectra with standard Fast Fourier Transform techniques on the 3-dimensional grid, assuming periodicity, and in the case of the velocity spectrum, $P_v(k)$, we weighted the velocity variable by the square root of gas density, $v'=\rho^{1/2} v$, so that the kinetic and magnetic spectrum, $P_B(k)$, have the same units. 
$P_v(k)$ is  indistinguishable in the six runs, indicating that differences in the magnetic field at the level explored in this work cannot affect the overall gas kinematics in any important way. On the other hand, the evolution of $P_B(k)$ shows substantial differences on all scales for $z \geq 1.0$, while at lower redshift the spectra overlaps on scales $\leq 10-20$ $\rm Mpc$, as a result of the dynamo amplification. The similarity between models at small spatial scales is driven by our subgrid dynamo model. Since the small-scale dynamo  erases any topological memory of initial fields in the real Universe \citep[e.g.][]{cho14,2015MNRAS.453.3999M,va18mhd}, all models look alike. Already from this test, it is clear that the differences between models at low redshift will be minute and difficult to observe, because most of the difference is contributed by very large spatial scales, $\gg 10$ $\rm Mpc$, which are characterised by low densities and large angular scales in the sky. 

\subsubsection{Properties of galaxy clusters}

We first examine the properties of our simulated population of galaxy clusters and groups, in order to check whether significant differences between extreme variations in the primordial seed fields can already be detected. 

The typical difference between the magnetic field topology within and around the most massive halo for all investigated models is shown in Fig.~\ref{fig:zoom}.

We identified all massive self-gravitating halos with a standard analysis of the spherical (gas+DM) overdensity \citep[e.g.][]{gpm98b}, and extracted the total mass and average (volume weighted) magnetic field strength  within the radius enclosing a matter density $200$ times the cosmological critical density ($R_{\rm 200}$). 

\begin{figure}
\includegraphics[width=0.45\textwidth]{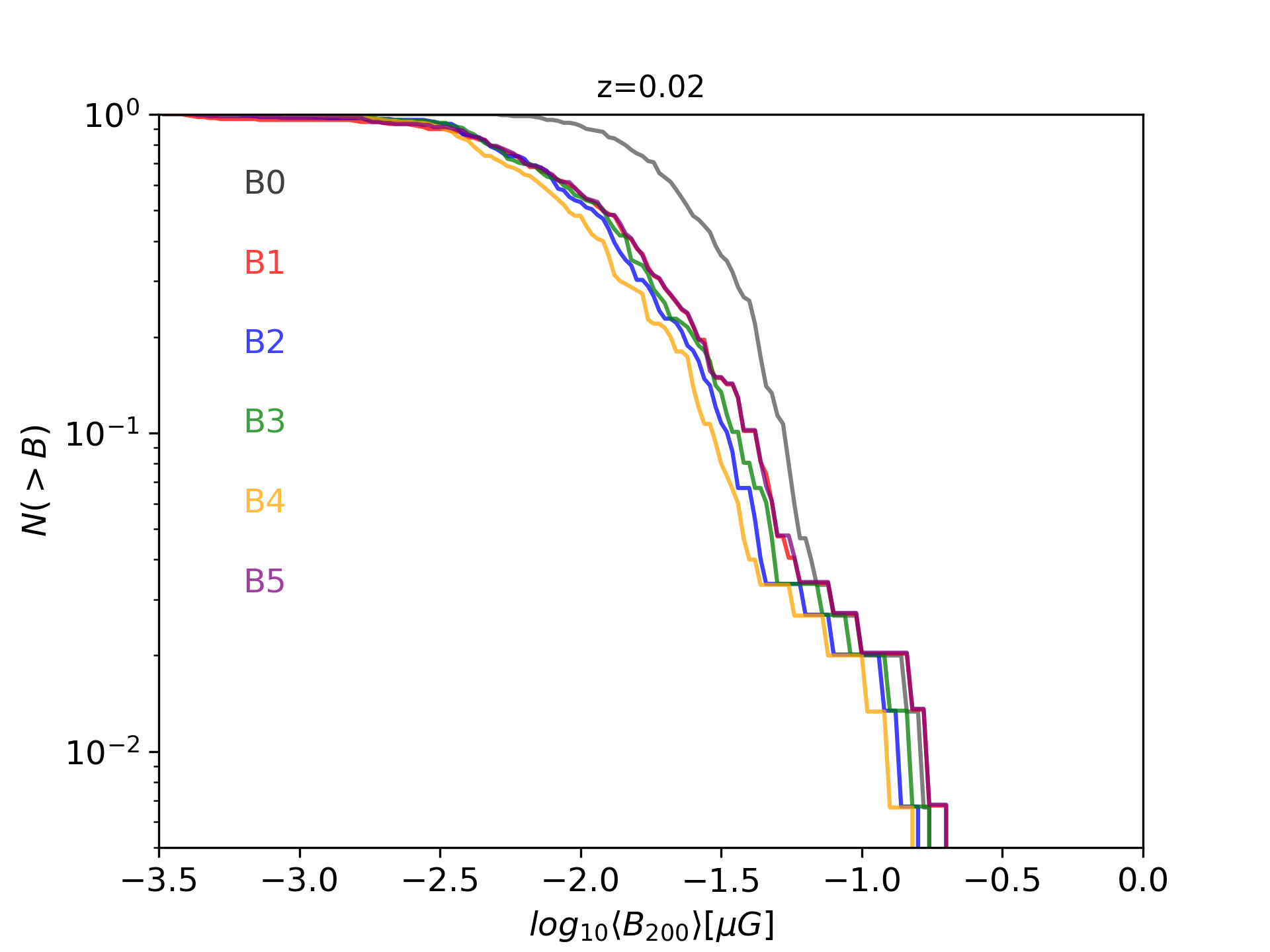}
\includegraphics[width=0.45\textwidth]{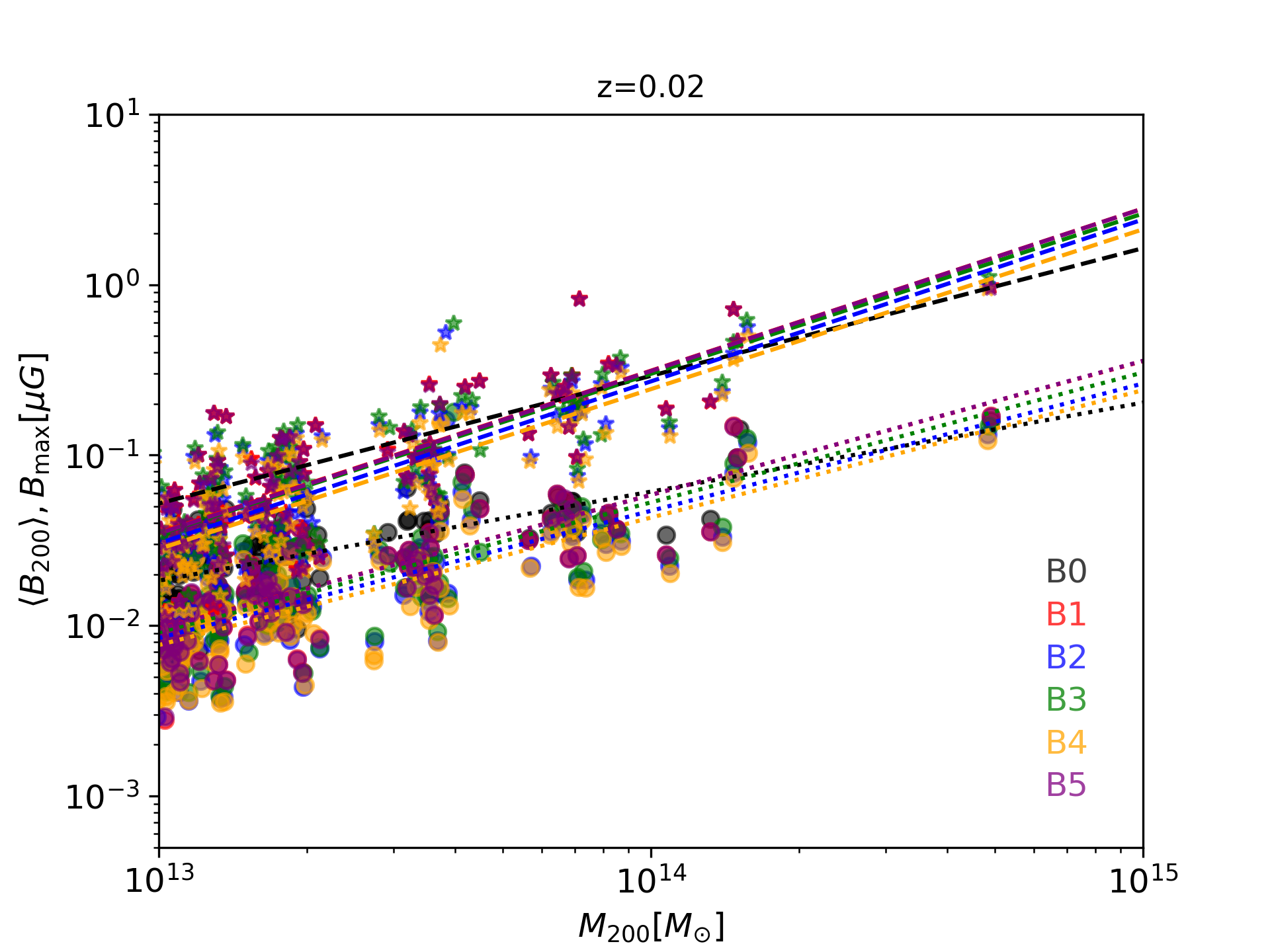}
\caption{Top: 
cumulative mean magnetic field strength (volume averaged within $R_{\rm 200}$) for the same halos. Bottom: scatter plot relating $M_{\rm 200}$ and the average ($\langle B_{\rm 200} \rangle$) or the maximum ($B_{\rm max}$) magnetic fields for the same clusters. The additional dashed/dotted lines gives the best fits of the two relations. }
\label{fig:mfunc}
\end{figure}

In principle, the presence of a large magnetic tension in the cosmological initial conditions may induce a change in the late evolution of the clustering properties of galaxy clusters, of their mass function as well as of their thermal scaling relations \citep[e.g.][]{do99,2001A&A...369...36D,2013ApJ...770...47K,2020arXiv200505401S}. 
However, the top panel of Fig.~\ref{fig:mfunc} shows that the mass function of halos identified in all simulations at $z=0.02$ are basically indistinguishable,  across more than two orders of magnitude in mass, down to $\approx 10^{12} M_{\odot}$. 
Moreover, the scaling relations between the cluster mass and the average gas temperature within $R_{\rm 200}$ show no differences (not shown). 

On the other hand, larger differences are measured for the average (volume weighted) magnetic field strength within $R_{\rm 200}$,  $\langle B_{\rm 200} \rangle$, between the same objects. The best-fit relations suggest that the differences are on average of order $\leq 10-20\%$ between extreme models. This makes the detection of any difference on these small scales extremely challenging, because in observations the global cluster magnetic field can be either estimated through the (sparse) sampling of Faraday Rotation from background polarised sources, and/or from the modelling of diffuse radio emission originating from shocks or turbulence \citep[e.g.][]{bj14,2019SSRv..215...16V}. 
 Therefore, we consider any differences of the magnetic field distribution inside clusters or groups of galaxies likely too weak to be detectable by observations. 

Unlike in \citet{2013ApJ...770...47K}, we consider only the stochastic background of primordial magnetic fields and do not account for the effect of magnetized perturbations on the matter power spectrum at small scales.
For this reason, the effects of different models on the final distribution of cluster masses might be underestimated. However, even in the more self-consistent (analytical) work by \citet{2013ApJ...770...47K}, significant effects of primordial magnetic fields on the halo mass function are confined to $\leq 10^{12} M_{\odot}$ halos for most models, i.e. in a range of masses which cannot be properly resolved by our runs here. In particular, \citet{2013ApJ...770...47K} reported that the number of small mass objects ($M \sim 10^4 M_{\odot}$) in the most magnetised cases can be reduced by a factor $\sim 10^2$ compared to the unmagnetised case number, while an excess in the number halos with respect to the unmagnetised case is found for  objects with masses of $\sim 10^{10} M_{\odot}$.

\begin{figure*}
\includegraphics[width=0.99\textwidth]{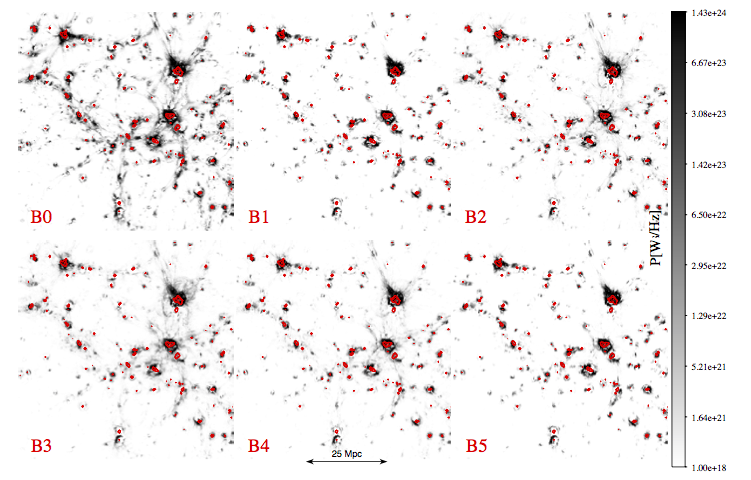}
\caption{Synchrotron radio emission ($\nu=100$ MHz) from cosmic shocks along the entire 100 Mpc, and projected DM density (red contours) for all our models at $z=0$.} 
\label{fig:Pradio}
\end{figure*}

\begin{figure}
\includegraphics[width=0.49\textwidth]{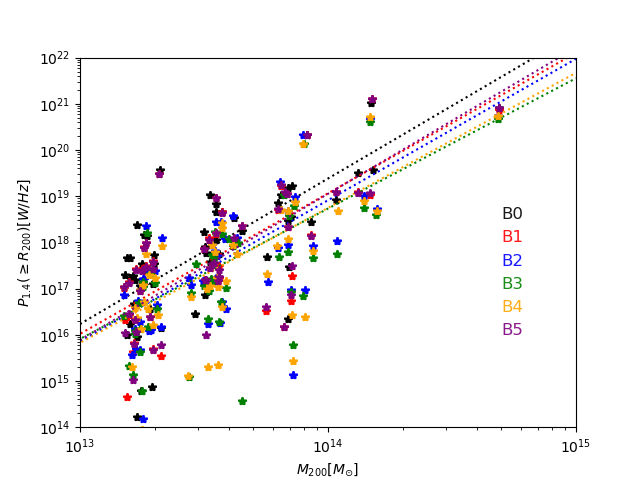}
\includegraphics[width=0.49\textwidth]{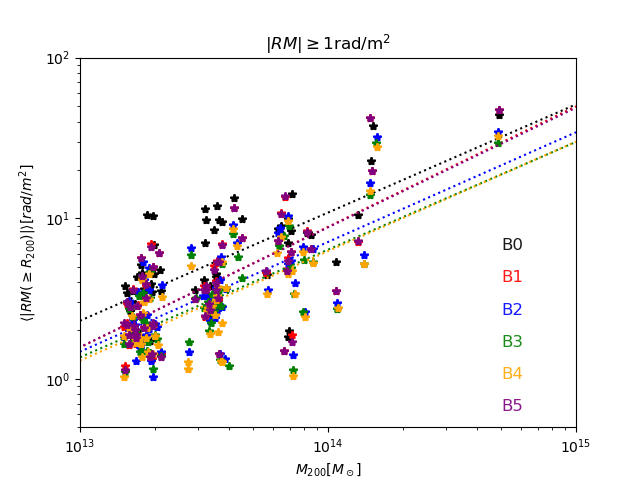}
\caption{Top: radio power from shocks at $1.4$ GHz from peripheral,  $R_{\rm 200} \leq r \leq ~2R_{\rm 200}$ regions  vs $M_{\rm 200}$ at $z=0.02$ for the same clusters in all runs.  The dotted lines show the best fit relations for each model. Bottom: average $|\rm RM|$ in the range $R_{\rm 200} \leq r \leq ~2R_{\rm 200}$, considering a (future) $|\rm RM| \geq 1 ~\rm rad/m^2$ sensitivity. The dotted lines show the best fit relations for each model.}
\label{fig:relics}
\end{figure}

\begin{figure*}
\includegraphics[width=0.99\textwidth,height=0.38\textwidth]{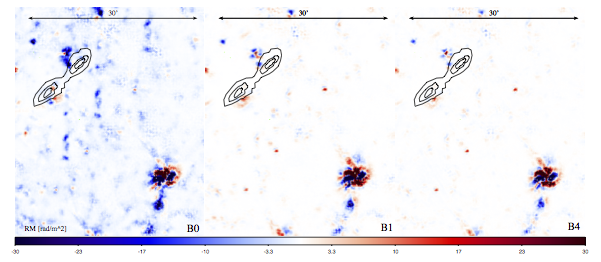}
\caption{Integrated Faraday rotation for a $\sim 30' \times 50'$ field of view with depth until $z=0.5$, for models B0, B1 and B4. The maps are convolved for a $20"$ resolution beam. The simple sketch of an FRII radio galaxy is added to refer to our analysis discussed in Sec.3.2.1.}
\label{fig:deltaRM_map}
\end{figure*}  

\begin{figure}
\includegraphics[width=0.49\textwidth,height=0.35\textwidth]{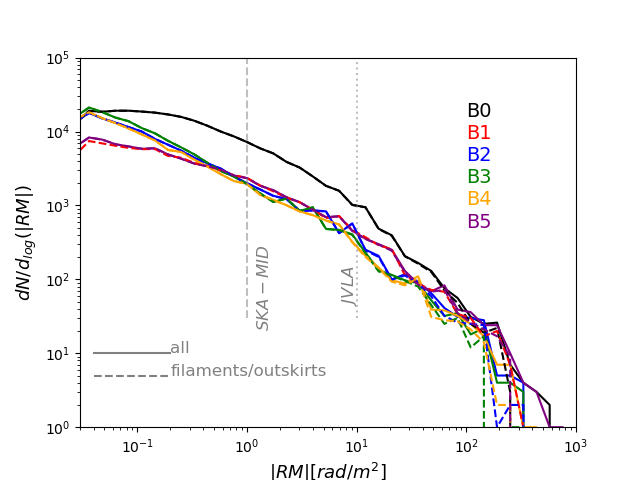}

\includegraphics[width=0.49\textwidth,height=0.35\textwidth]{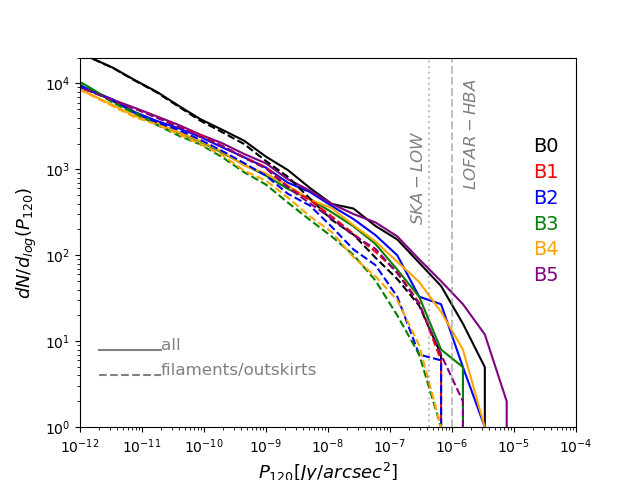}
\caption{Top: distribution functions of Faraday Rotation for all models and a $2^{\circ} \times 2^{\circ}$ field of view up to $z=0.8$. Bottom: distribution function of synchrotron radio emission at 1.4 GHz for all models and a $2^{\circ} \times 2^{\circ}$ field of view up to $z=0.8$.}
\label{fig:pdf_RM}
\end{figure}

\subsection{Observable properties}

We compute the  {\it Faraday Rotation Measure} (RM) across the full length  of the simulated box, by integrating for each 1-dimensional beam of cells the quantity: 

    \begin{equation}
  \frac{\rm R}{[\rm rad/m^2]}=812 \int \frac{B_{\rm ||}}{\rm [\mu G]} \cdot \frac{n_e}{\rm [cm^3]} \frac{dl}{[\rm kpc]}\frac{1}{(1+z)^2},
\end{equation}

in which $B_{\rm ||}$ denotes the component of the magnetic field parallel to the line of sight (LOS), $z$ is the redshift of each cell,  $n_e$ is the physical electron density of cells, assuming a primordial chemical composition of gas matter everywhere in the volume ($\mu=0.59$). RM thus 
corresponds to the Faraday Rotation experienced by sources of  linearly polarised radio emission, located in the background of our LOS. 

To compute the  diffuse {\it synchrotron radio emission} from the cosmic web, we limit here to the contribution from cosmic shocks \citep[e.g.][]{va15radio} and assume that relativistic electrons get accelerated by DSA. We closely follow the formalism by  \citet{hb07}, and compute the synchrotron emission for each shocked cell (identified  with a velocity-based approach as in \citealt[][]{va09shocks}) as the convolution of the several power-law distributions of electrons that overlap in the cooling region downstream of each shock: 

\begin{eqnarray}
\rm \frac{P(\nu)}{[\rm erg/s/Hz]}=6.4 \cdot 10^{34} \cdot \int  \frac{\xi_e(\mathcal{M}) \cdot A}{\rm Mpc^2} ~\frac{n_e}{10^{-4}\rm cm^{-3}} \nonumber\\ 
~(\frac{T}{7 \rm keV})^{3/2} (\frac{\nu}{\rm GHz})^{-s/2} \cdot \frac {B^{1+s/2}}{B_{\rm CMB}^2+B^2} \frac{1}{(1+z)^2}dV,
\label{eq:hb}
\end{eqnarray}

in which $A$ is the shock surface, $T$ is the cell temperature, $s$ is the energy spectrum of accelerated particles, defined as $s=2(\mathcal{M}^2+1)/(\mathcal{M}^2-1)$, $B$ is the magnetic field (in $\rm \mu G$ and  $B_{\rm CMB}$ is the CMB-equivalent magnetic field. $\nu$ is the observing frequency, $dV$ is the cell volume and  $\xi_e(\mathcal{M})$ is the acceleration efficiency of electrons, which is $\leq 10^{-4}$ for weak shocks ($\mathcal{M} \leq 3-4$) and it reaches  $\sim 10^{-2}$ for strong, $\mathcal{M} \gg 10$ shocks. 


We give an example of the distribution of synchrotron radio emission (with the projected Dark Matter density overlaid) emitted from the shocked cosmic web for all our models in Fig.~\ref{fig:Pradio}.

The panels in Fig.~\ref{fig:relics} give the distribution of RM and of radio power for all galaxy clusters in our volume, limited to their peripheral $\geq \rm R_{\rm 200}$ regions, which are the typical location of peripheral radio relics \citep[][]{2019SSRv..215...16V}.

The best-fit relations for the relation between the radio emission or the Faraday Rotation and $M_{\rm 200}$ of halos have quite similar slopes, while their differences in normalisation follow from the different trends in the average magnetic field found in Fig.~\ref{fig:mfunc}. Interestingly, this suggests that already outside of $R_{\rm 200}$ there is a residual difference in radio observables, that is not entirely erased by the action of the small-scale dynamo amplification. A higher value of RM and synchrotron emission is expected for the simple B0 model, consistent with the systematically higher average fields found in the previous Section. However, the scatter present in both quantities within all mass bins is much larger than the difference between models. Because of this, only with very large surveys of clusters it might be possible to reduce the effect of cosmic variance, which are instead a big limitation with samples of $\leq 50$ objects as the one simulated here. 

\subsubsection{Cosmological light cones}

In the remainder of the paper, we consider the integrated synchrotron radio emission or Rotation Measure from light cones which are longer than our simulated volume, in order to produce a more realistic match with what observations can do.
In detail, we used stacked $10$ different redshift snapshots for each of our model resimulations, by extracting a corresponding cone within a fixed aperture ($\theta=2^{\circ}$). The light cone was build by adding the information within each volume at  regular intervals of $100 \rm ~Mpc$ (comoving) along the line of sight, and by introducing random shifts across the plane of the sky to avoid repeating patterns, until producing a $0.02 \leq z \leq 0.8$ light cone. An example of the integrated Faraday Rotation for a light cone in the three reference models B0, B1 and B4 is shown in Fig.~\ref{fig:deltaRM_map}.

We show the resulting distributions of 
synchrotron radio emission (at $\nu=1.4 \rm ~GHz$), and of Faraday Rotation for background polarised sources in  Fig.~\ref{fig:pdf_RM}. We additionally mark the regions traced by "filaments" based on the projected mass-weighted gas temperature, using a fiducial threshold of $T_{\rm fila}=10^{7} \rm K$ to disentangle the signal produced by halos from the one produced by more rarefied regions. 

Small, but non negligible differences, appear between different models, albeit in a regime which is challenging to observe with current instruments. There is a level of difference from filaments in the cosmic web (dotted lines), which seems to be at the edge of detectability both in Faraday Rotation and in synchrotron emission. 
It is also worth noticing  that there is a systematic difference in Faraday Rotation between a simplistic uniform primordial magnetic field model (B0) and all other models. The B0 model yields a $\sim 3-5$ higher Faraday Rotation for filaments, compared to all other models, consistent with our previous results on clusters, and following from the fact that the rather laminar magnetic field lines in the bulk of cosmic volume do produce a systematic Faraday Rotation of background sources.

\begin{figure}
\includegraphics[width=0.45\textwidth,height=0.34\textwidth]{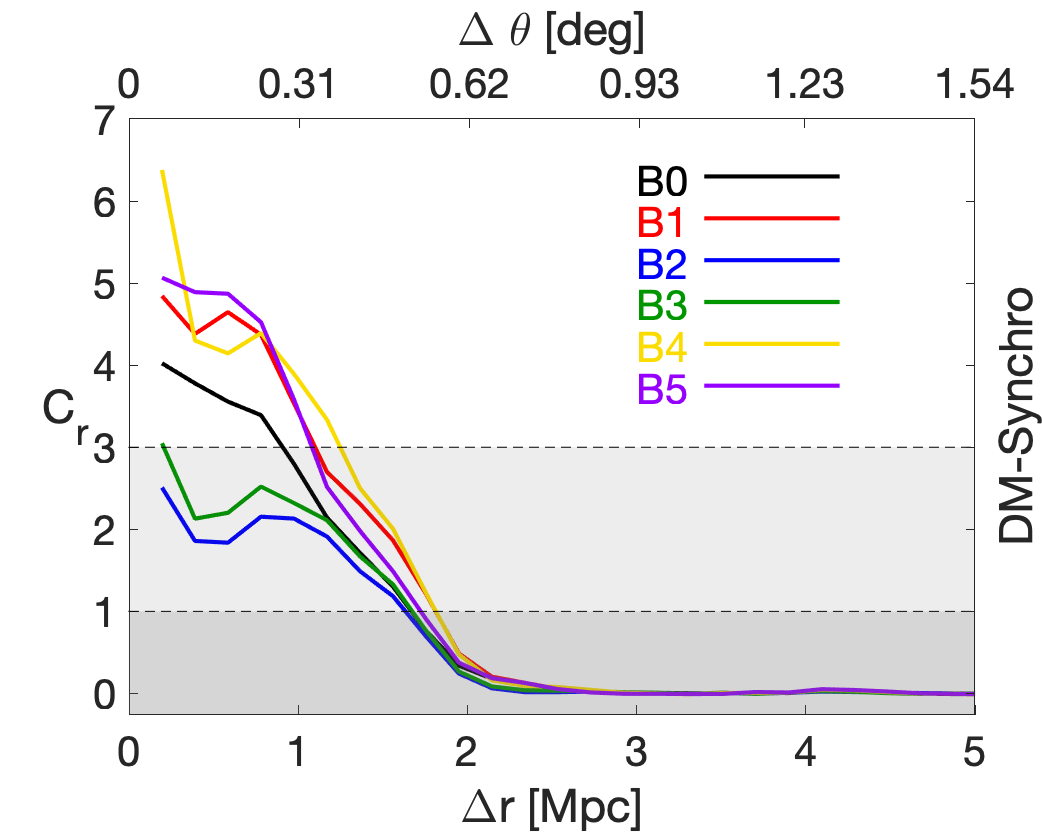}
\caption{Simulated cross-correlation between the projected dark matter distribution (simulating galaxies) and the synchrotron emission detectable by MWA, using the observing configuration by \citet{vern19}. The cross-correlation is normalised to the corresponding null model and the horizontal lines mark the 1 and 3 $\sigma$ detection level.}
\label{fig:cc}
\end{figure}  

\subsubsection{Comparison to available radio observables }

A few authors have claimed a derivation of upper limits on the average magnetisation of the cosmic web, based on the statistical analysis of radio surveys \citep[e.g.][]{vern17,vern19,os20}. Our simulations can already be used for a first tentative comparison with recent work, albeit within the caveat posed by their limited resolution, which are discussed in detail in Sec.4 and in the Appendix.

First, we attempt a qualitative comparison with the recent 
results by \citet{vern17}, who have cross-correlated the MWA Phase I radio observations and the large-scale  distribution of galaxies observed with the WISE+2MASS galaxy survey, for a  $22^\circ \times 22^\circ$ field of view. \citet{vern17} reported no significant detection of cross-correlation between galaxies and radio emission  on $\geq 20'$ scales, and used this information to derive upper limits on the typical magnetisation of filaments of the cosmic web. In \citet{gv20} we introduced a procedure to simulate the  the cross-correlation between the diffuse radio emission at 180 MHz and the projected galaxy distribution in the simulation. 
In particular, we set the threshold of projected DM density to $\rho_{th} = 6 \cdot 10^{-29} \rm ~g/cm^3$ in order to produce a number of galaxies (i.e. dark matter halos) compatible with the sensitivity of the WISE IR survey, yielding $\sim 10$ galaxies per square degree for $z \leq 0.07$  \citep[][]{vern17}. 
For the radio emission, we considered the MWA Phase I sensitivity and resolution beam as in \citet{vern17}, by  convolving our radio sky model for a $\theta \approx 2.9'$ resolution beam and considering a (spatially uniform) noise level of $0.96 \rm ~mJy/beam \approx 0.028 \rm ~\mu Jy/arcsec^2$ at $180$ MHz.

Following \citet{gv20}, the normalised correlation matrix $C$ between 2D $N\times M$ pixels images $A$ and $B$ is evaluated through:

\begin{equation}
    C(k,l) = {1\over NM}\sum_{j=0}^{N-1} \sum_{i=0}^{M-1} {(A(i,j)-\bar A)(B(i+k,j+l)-\bar B) \over 
               \sigma_A \sigma_B},
\end{equation}
where $\bar A$ and $\bar B$ are the mean values of the two images and $\sigma_A$ and $\sigma_B$ are their standard deviation, while the  indices of the correlation matrix $C$ give the shift (displacement) of the two images. 
The significance of the cross-correlation is evaluated against the case of null correlation, for which $C$ is computed between $A$ and $B$ from uncorrelated $50 \times 50 \rm ~Mpc^2$ sub-tiles extracted from the main simulation. 

The results are shown in Fig.~\ref{fig:cc}; 
they cannot be readily compared with the cross-correlation values given by \citet{vern17}, owing to the different approaches in estimating the noise level of the cross-correlations. Therefore, our synthetic observation can only qualitatively address which model seems to be more compatible with MWA observations.
 While all models give a significant cross-correlation out to $\sim 40-50'$, the amplitude of the correlation in the B1, B4 and B5 models (and, to a smaller extent, over the homogeneous B0 model) seems to be too large to have been missed by MWA observations. This potentially suggests already that a $1 \rm ~nG$ initial field  (resulting into a typical magnetisation of filaments of $10-100$ nG as shown in \citealt{2019MNRAS.486..981G}) is too large to be compatible with the lack of detection reported by \citet{vern17}. Interestingly, intermediate models B2 and B3 present cross-correlation only at the $\leq 2 \sigma$ level and on slightly smaller angular scales. 

As a caveat, given the finite mass resolution of our simulations, we cannot properly form dwarf galaxies in voids (or in very poor environment, in general). Therefore, even if the number of galaxies is calibrated to be at the level of the galaxy distribution observed in WISE/MASS surveys, our spatial distribution is typically more clustered than in observations. In principle, this can decrease the cross-correlated signal coming from low density regions in our sample. With future work, we will employ more resolved simulations in order to address this issue.

\citet{os20} 
recently reported limits on the Faraday depth contribution of the magnetised cosmic web, by comparing the structure function of Faraday Rotation in random pairs of radio galaxies compared to physical radio galaxies (i.e.~doubled-lobed radio galaxies), on overlapping angular scales. 

We produced mock  structure functions of Faraday Rotation for our simulated deep light cones, following the same procedure as in  \citet[][]{os20}. 
We  integrated the RM along the line of sight down to $z=0.7$, and computed the statistic of $\langle (\Delta RM)^2\rangle$ for physical pairs of pixels, by placing pairs of sources at regular intervals of $100 \rm ~Mpc$ (comoving) along the line of sight (i.e. at the end of each of the comoving volumes used to produce the stacking sequence of Faraday Rotation). In detail, we first randomly drew $20,000$ sources (with $|\rm RM| \geq 0.03 \rm ~rad/m^2$ as in the LOFAR observation, for $10$ evenly spaced redshift bins, and  calculated $(\Delta RM(\Delta \theta)^2)$  at each redshift. The assumed angular resolution for the light-cone of Faraday Rotation is the same of \citet{os20} LOFAR-HBA observation, i.e. $\approx 20"$.  The final distribution  of $(\Delta RM(\Delta \theta)^2)$ can be compared with  recent LOFAR-HBA observations by \citet{os20}, by weighting each structure function in redshift bins for the (likely) redshift distribution function of sources, derived in \citet{vern19}. 

\begin{figure}
\includegraphics[width=0.49\textwidth]{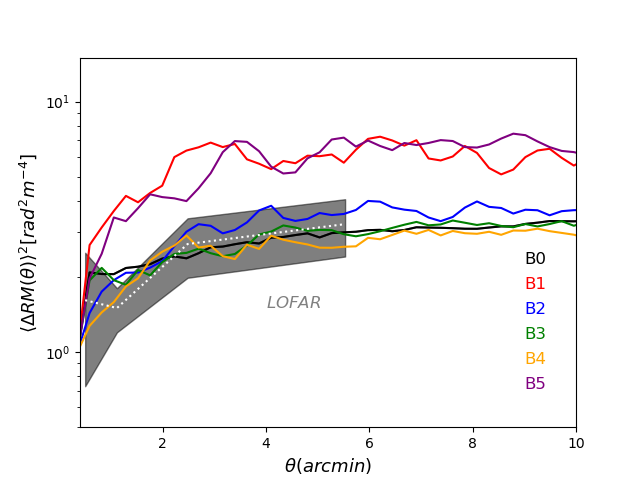}
\caption{Simulated distribution of $(\Delta RM(\Delta \theta)^2)$ as a function of angular separation for all our models. The grey area shows the LOFAR 144 MHz data by \citet{os20}, with noise power from measurement errors subtracted, for physical pairs of radio sources only.}
\label{fig:sim_rm}
\end{figure}

Fig.~\ref{fig:sim_rm} gives the simulated distribution of $(\Delta RM(\Delta \theta)^2)$ as a function of angular separation for all our models, compared with real LOFAR-HBA data (grey area).
At face values, the structure functions of models B1 and B5 appear strongly disfavoured by the comparison with observations, while all other models remain compatible with LOFAR data, even if it is non trivial to evaluate the uncertainties in the signal from simulations, owing to the limited volume scanned here as well as to the  different selection function of sources in redshift applied to simulated data.  
The trend for  B1 and B5 is in striking good agreement with the previous test on the cross-correlation of synchrotron emission and the galaxy distribution, in the sense that these two models already likely to ruled out by radio observations.
However, our limited spatial resolution is likely to underestimate the Faraday structure below $\leq 200 \rm ~kpc$, with a stronger effect than in the case of synchrotron emission, because Faraday Rotation is more sensitive to the (unresolved) small-scale topology of magnetic and density perturbations. 
We comment on this issue more in detail in Sec. 4 and in the Appendix. 
Moreover, it is important to notice that the above structure functions still have an  unknown contribution from the Rotation Measure variance of the Milky Way. In case this is dominant on $\leq 10'$ scales, the observational constraints from $(\Delta RM(\Delta \theta)^2)$ will decrease to lower values, possibly putting tension on all models investigated here.

\subsubsection{The deflection of Ultra-High-Energy Cosmic Rays}
Finally, we investigated whether the propagation of Ultra-High-Energy Cosmic Rays across cosmic distances is differently affected by our different magnetic field models.
Cosmic rays with energies $\geq 10^{18} ~\rm EeV$ are indeed believed to have a predominantly extra-galactic origin, and their propagation towards Earth must be deflected by the Lorentz force by extragalactic magnetic fields \citep[e.g.][]{Sigl:2003ay,2005JCAP...01..009D}. Large deflections hamper the possibility of  locating the real sources of UHECRs at the highest energies\citep[e.g.][]{2016arXiv161000944B}, yet the amplitude of these deflections can greatly vary depending on model assumptions, ranging from $\leq 1^{\circ}$ to $\sim 30^{\circ}$ \citep[e.g.][]{2005JCAP...01..009D, 2014APh....53..120B,2014JCAP...11..031A}.  We recently studied   the propagation of UHECRs in {\enzo} simulations combined to the  CRPropa code \citep[][]{2016JCAP...05..038A}, finding that the observed level of isotropy of $\geq 10^{18} \rm eV$ UHECRs rules out primordial models with a uniform magnetic field larger than $\sim 10~ \rm nG$ \citep[][]{hack16,hack19}.

The typical deflection angle, $\theta$, of a cosmic ray of energy $E$ and charge $Z$ propagating through the extragalactic magnetic field can be simply estimated as

\begin{equation}
\theta(E)=0.8 \rm ~Z~\big (\frac {E}{10^{20}\rm eV}\big)^{-1}\cdot \big(\frac{r}{10 \rm Mpc}\big)^{1/2} \cdot \big(\frac{\lambda_B}{\rm Mpc} \big )^{1/2} \cdot \frac{B}{\rm nG}
\end{equation}

where $\lambda$ is the typical coherence length of magnetic fields (assumed in this case to be constant across the cosmic volume, and fixed to $\lambda_B = 1 ~\rm Mpc$ here for simplicity), and $r$ is the propagation length \citep[e.g.][]{Sigl:2003ay}.

\begin{figure*}
\includegraphics[width=0.329\textwidth]{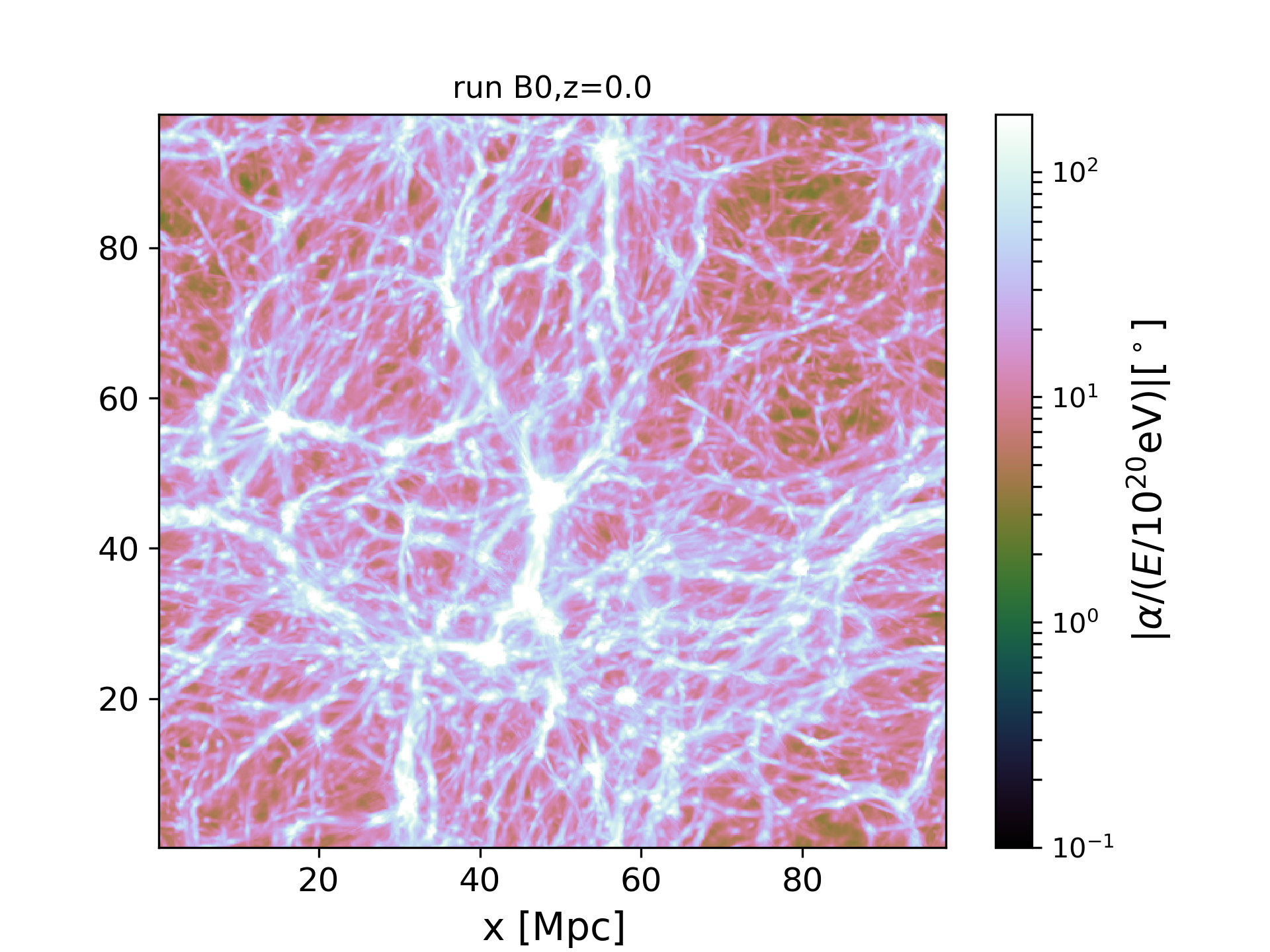}
\includegraphics[width=0.329\textwidth]{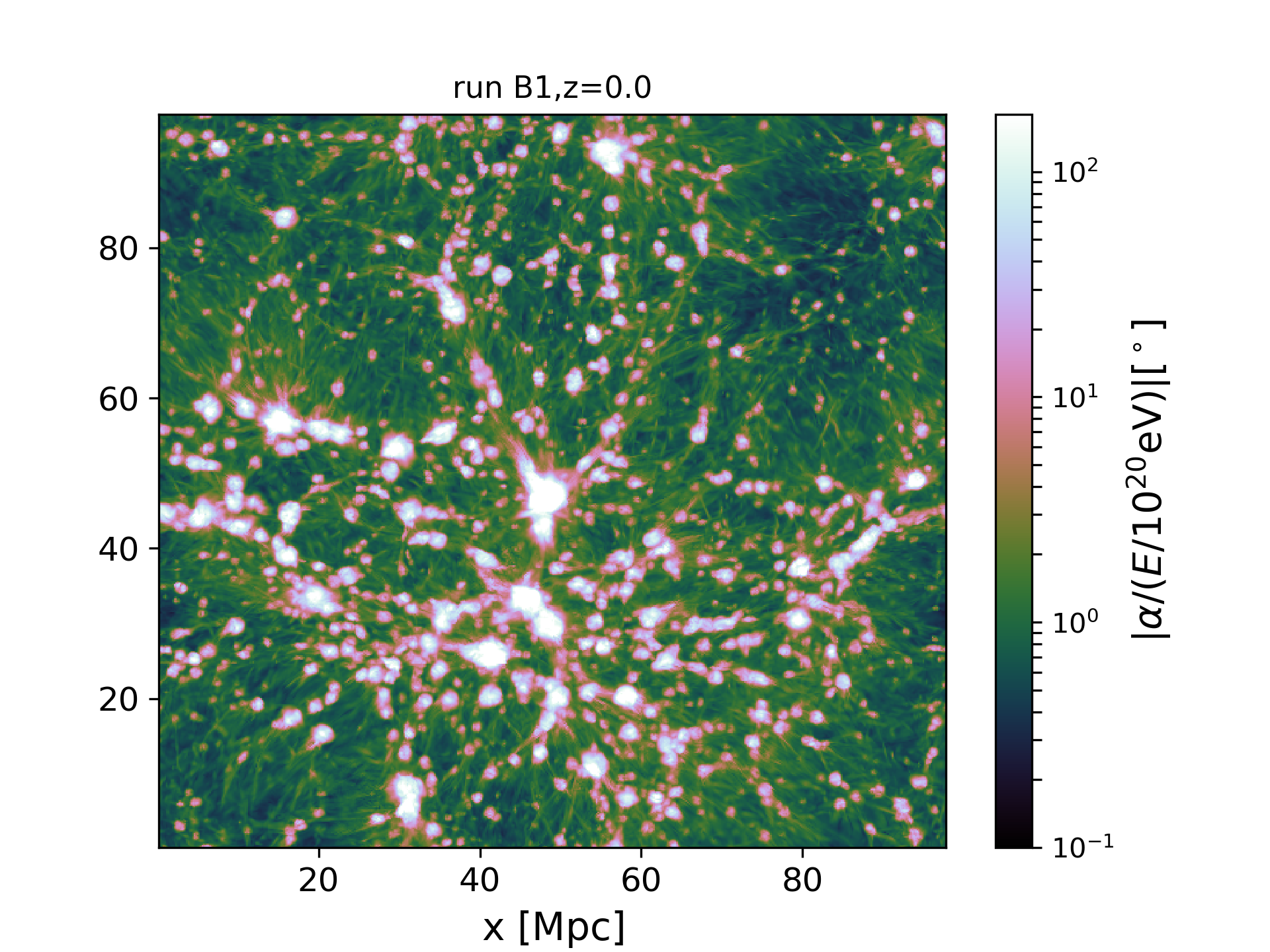}
\includegraphics[width=0.329\textwidth]{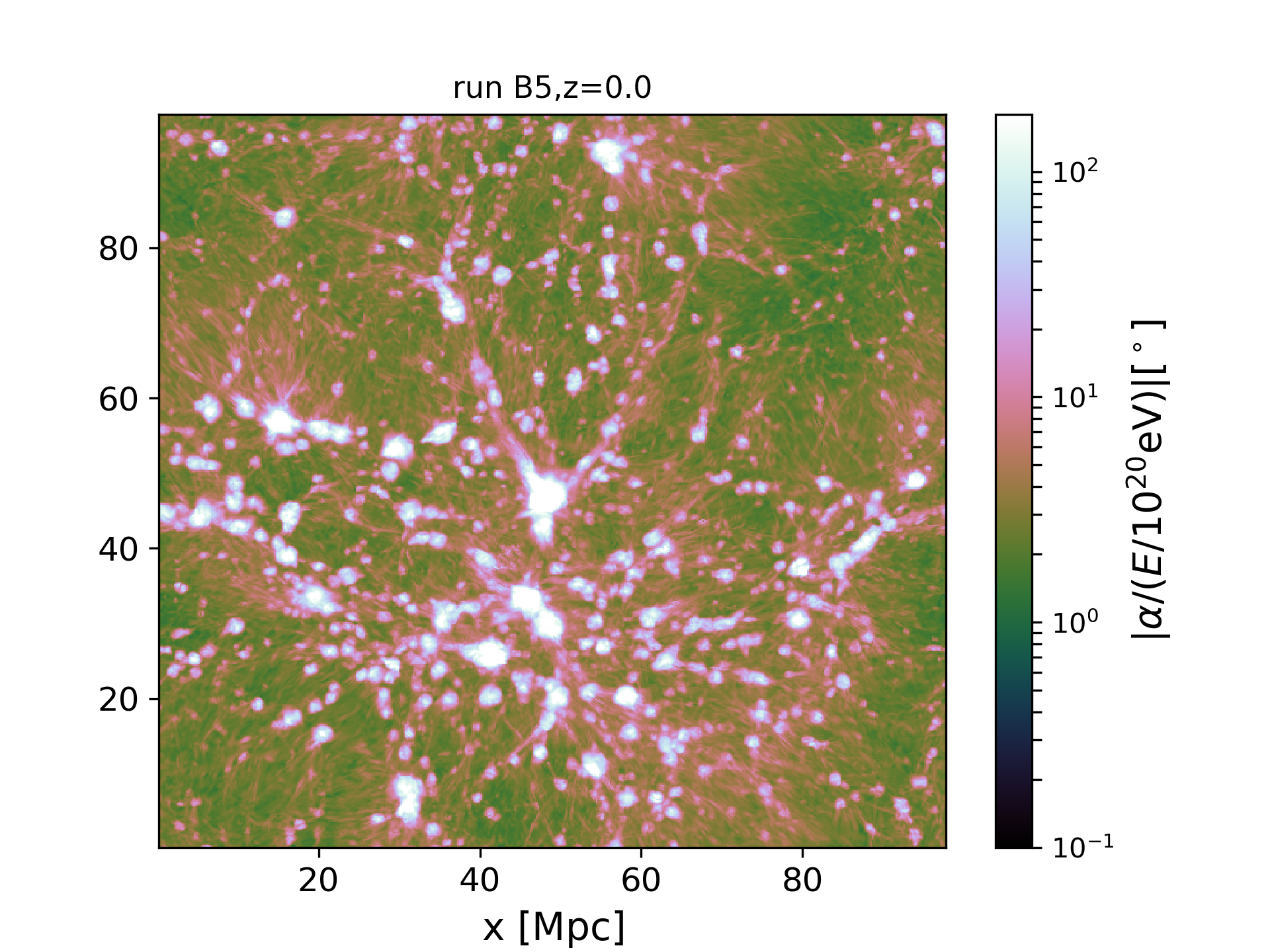}
\caption{Deflection angle for $E=10^{20} \rm eV$ protons, averaged along a 100 Mpc line of sight in models B0, B1 and B5.}
\label{fig:UHECR}
\end{figure*}

The maps in Fig.~\ref{fig:UHECR} give a visual impression of the average deflection angle computed from the above formula, for runs B0, B1 and B4 at $z=0.02$.
Very large ($\theta \gg 10^{\circ}$) deflection angles are expected for UHECR protons crossing the virial regions of halos, due to the $\geq 0.1 ~\rm \mu G$ field developed there. In the case of the uniform B0 model however the deflection is significant in filaments too, and far from negligible for the entire projected sky ($\theta \sim 10^{\circ}$). Deflections are still non negligible in the B5 model ($\theta \sim 2^{\circ}$), while in the model B1 the deflection is almost negligible ($\theta \leq 1^\circ$) for most of the projected area, once halos are excluded.

\begin{figure*}
\includegraphics[width=0.32\textwidth]{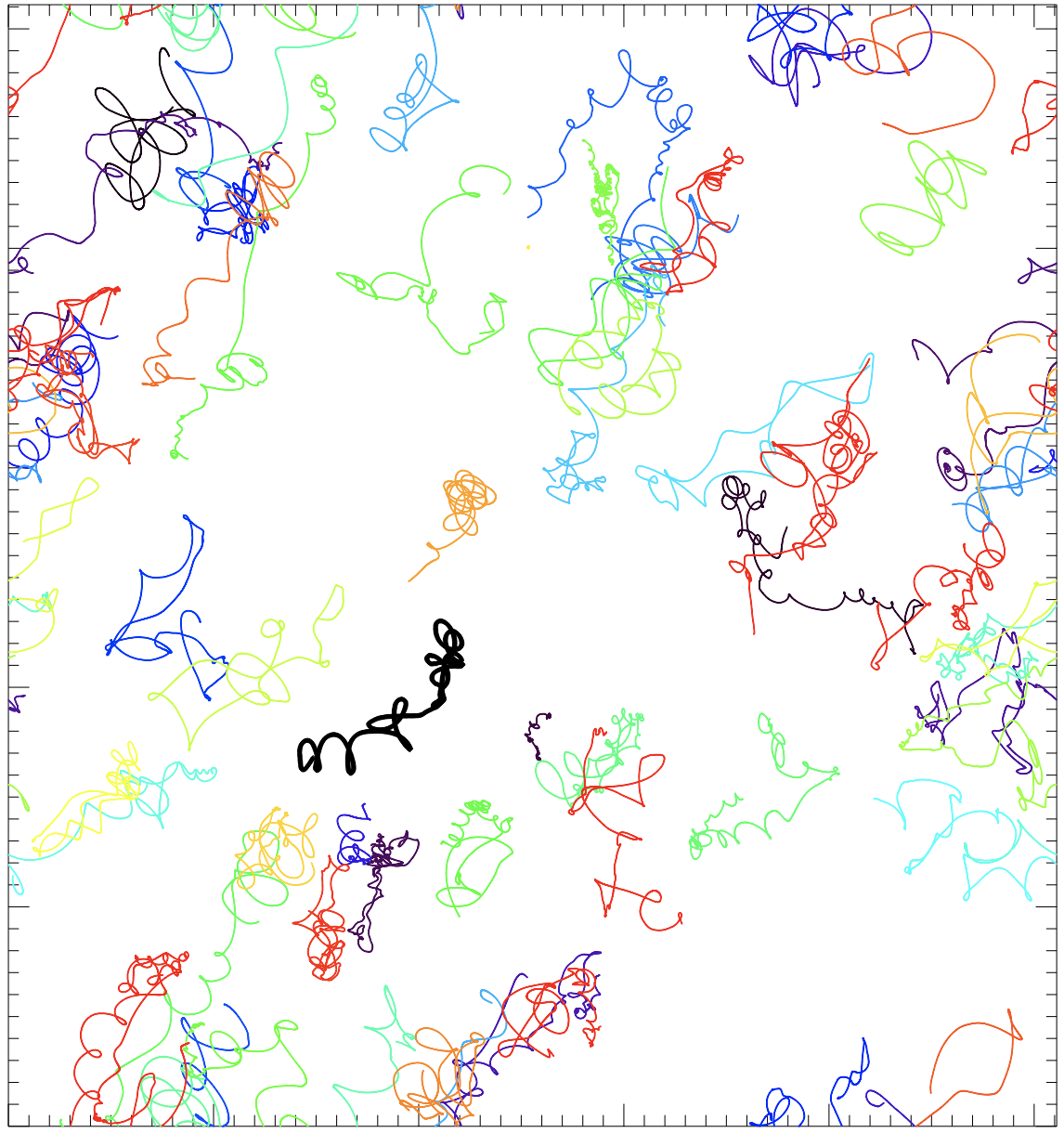}
\includegraphics[width=0.32\textwidth]{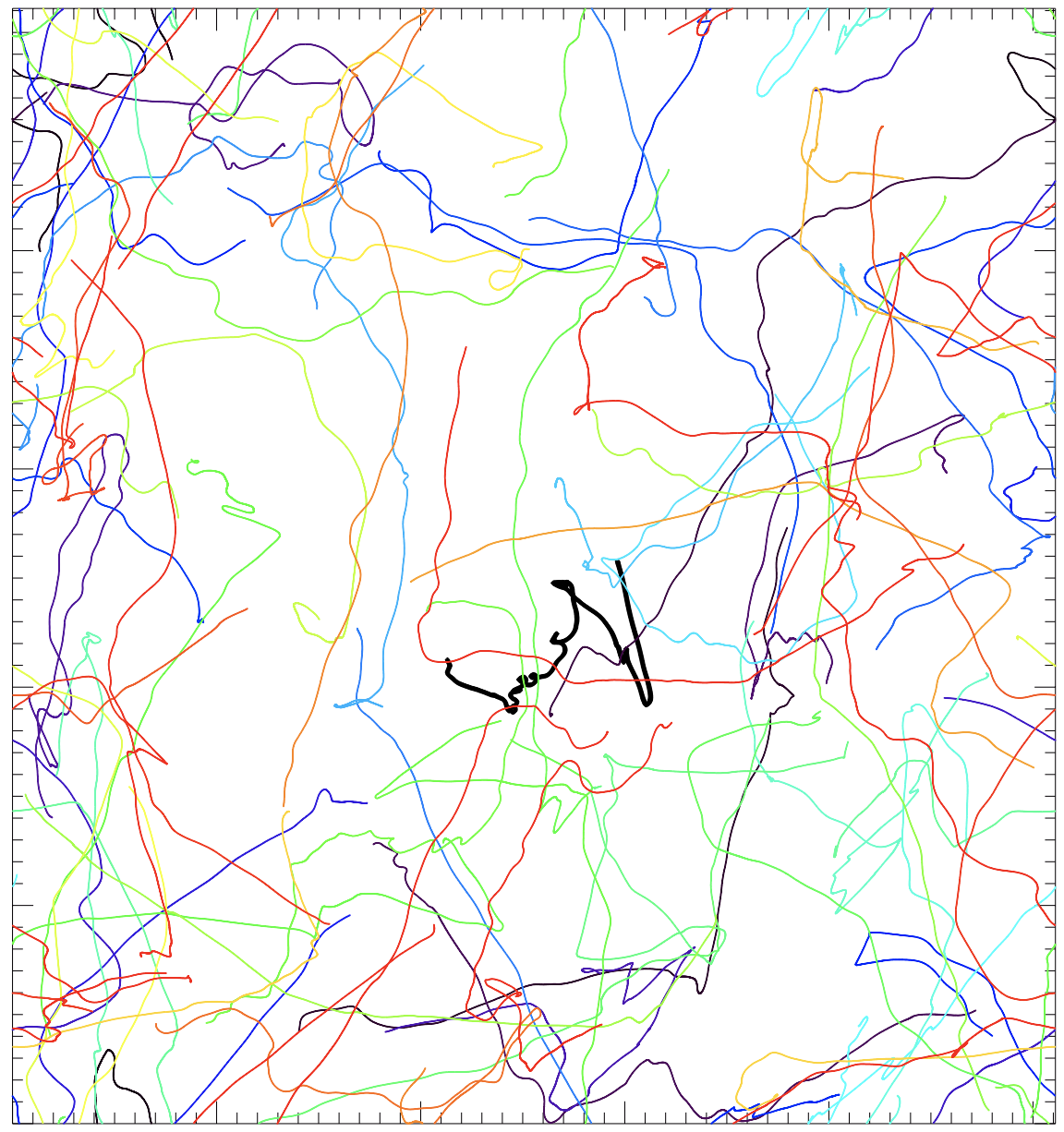}
\includegraphics[width=0.32\textwidth]{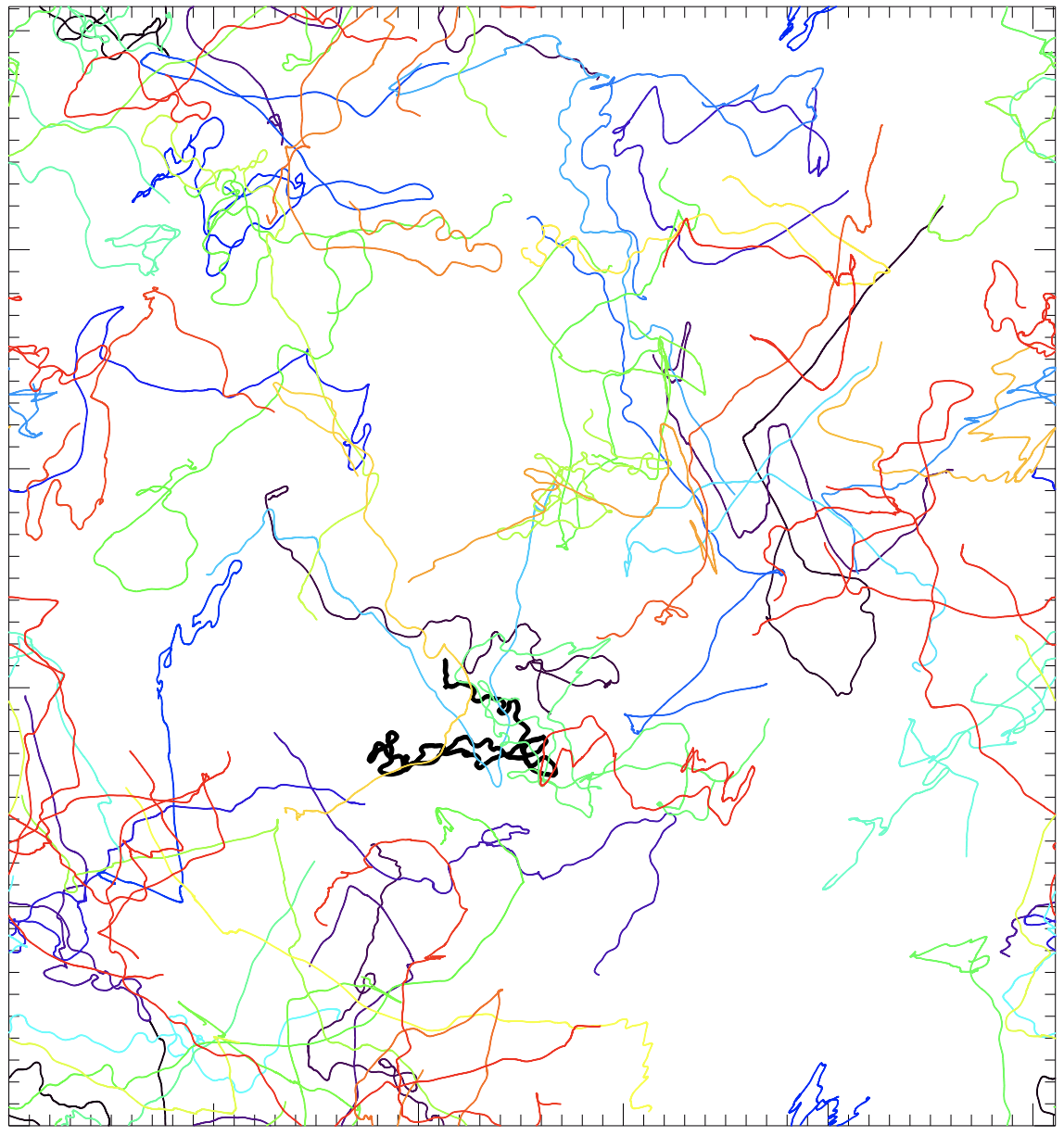}
\caption{Trajectories of $E=10^{19} \rm eV$ protons crossing our entire $100^3 \rm Mpc^3$ volume in models B0, B1 and B5. Each cosmic ray has been randomly injected at the opposite side of the box with respect to the observer.}
\label{fig:UHECR2}
\end{figure*}

\begin{figure*}
\includegraphics[width=0.329\textwidth]{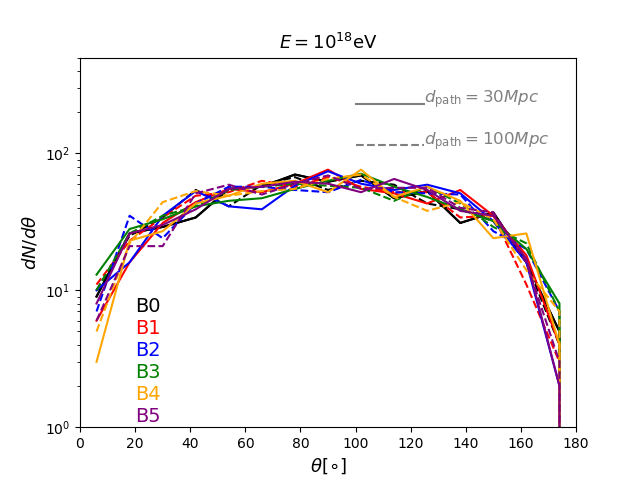}
\includegraphics[width=0.329\textwidth]{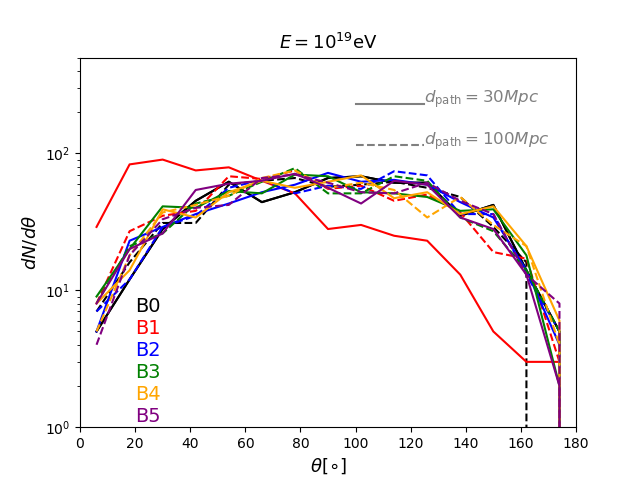}
\includegraphics[width=0.329\textwidth]{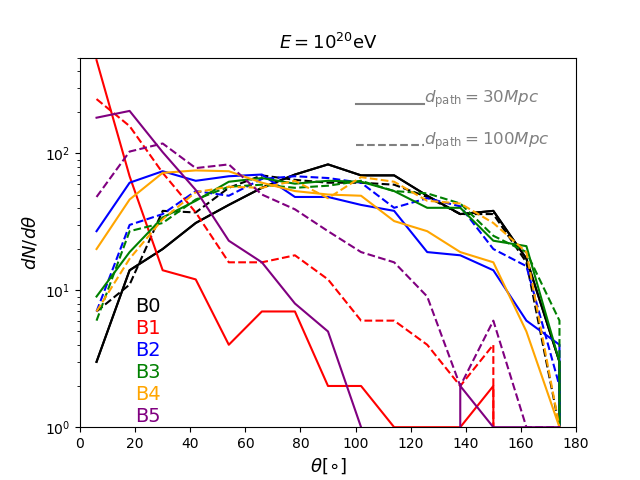}
\caption{Distribution of deflection angles for simulated UHECR protons with an initial energy  $E=10^{18} \rm eV$ (top) or $E=10^{19} \rm eV$ (centre) and $E=10^{20} \rm eV$ (bottom), injected with initial random positions and velocities in our simulated volumes. The different linestyles in both panels show the distribution of deflection angles after propagating for $30$ Mpc (solid) or $100$ Mpc (dashed).}
\label{fig:UHECR3}
\end{figure*}

A proper simulation of the effect of deflection by extragalactic magnetic fields requires to properly integrate the  trajectories of cosmic rays crossing our simulated volume. 
We thus simulated the propagation of UHECRs by integrating the trajectories of 2000 randomly distributed cosmic rays for each model, assuming a random initial orientation of their velocity (which is of course $|v|=c$) and we integrated their propagation through the simulated $100^3 ~\rm Mpc^3$ volume in all models. In detail, their propagation was integrated by computing the deflection by the Lorentz force with a kick-drift-kick second-order integration scheme \citep[e.g.][]{2010ApJ...710.1422R}, assuming that  all cosmic rays are protons ($Z=1$) with an initial energy of $10^{18} \rm eV$, $10^{19} \rm eV$  or $10^{20} \rm eV$, and neglecting all other loses of energy via other processes, which is a reasonable enough assumptions considering the relatively small propagation length considered here. The maps in Fig.
~\ref{fig:UHECR2} show the paths of a sample of protons through runs B0, B1 and B5, while in Fig.~\ref{fig:UHECR3} we give the distributions of deflection angles for the protons with the three possible initial energies, by computing $\theta$ as the angle between the initial injection direction and the final velocity vector, either after $30$ or $100$ $\rm Mpc$ of propagation length since their origin.
Consistent with the expectations of deflection angles, we observe a large difference in the typical propagation of UHECR protons in the B0, B1 and B5 case. The largest deflection is measured in the B0 scenario, in which the distribution of directions of UHECRs of $10^{20} \rm eV$ gets entirely randomized after propagating for $\sim 100$ $\rm Mpc$. For most of observers in this simulated Universe it will be thus entirely impossible to track the origin of received UHECRs, backwards to their original injection location. In most investigated models, already after $\sim 30$ Mpc of propagation the observed UHECR protons (for $E \leq 10^{20} \rm eV$) are so highly deflected, that that "UHECRs astronomy" would impossible. Only in the B1 and in the B5 models  $10^{20} \rm eV$ protons are not too much deflected after $30$ Mpc of propagation, and  their deflection also peaks at small angles after $100$ Mpc. 

This is consistent with previous results that investigated the propagation of UHECRs (also taking into account their energy losses) with {\enzo} simulation and a similarly large primordial field \citep[][]{hack16,hack19}.

In summary, we find that among the investigated models there is the theoretical possibility of detecting signatures of the topology of seed magnetic fields in voids, through the clustering of $10^{20} \rm eV$ events from very nearby $\leq 10 \rm ~Mpc$ sources. Unluckily, the statistics of such events is presently so small, that such analysis will be dominated by sample variance in the real case \citep[][]{hack19}.

\section{Discussion}
\subsection{Numerical \& physical limitations}

A few important limitations to our work, connected to the employed numerical scheme, must be taken into account in the interpretation of our results.  

Probably, the most important limitation of this set of simulation  is the adopted spatial resolution, which combined with our sub-grid dynamo model is enough to follow the evolution of magnetic fields and shocks on $\geq 200 \rm kpc$ scales  (as discussed in several previous work, e.g. \citealt{va14mhd,va14curie}), but likely not enough to simulate cell-by-cell fluctuations of RM, which are key to interpret the structure functions discussed in Sec. 3.2.2. 
We present in the Appendix the additional tests with a smaller series of runs, designed to test the dependence of our results on the adopted spatial resolution. As expected, the structure function for RM shows a significant, albeit non-dramatic, evolution with resolution. This presently makes it impossible to firmly establish which of our models for primordial magnetic fields can already be ruled out by comparison with the LOFAR data of \citet{os20}, and only future simulations at a higher resolution will be able to make a quantitative comparison.

We notice that RM fluctuations on scales smaller than the LOFAR synthesised beam ($20"$ i.e .$<150$~kpc at z=0.8) would cause wavelength-dependent depolarization \citep[e.g.][]{1966MNRAS.133...67B}. In this case, the emission from a population of sources that probe regions of high RM variance on small scales are likely to be missing from the LOFAR data at 144 MHz (i.e. completely depolarized), but would be present in data at a higher frequency, such as at 1.4 GHz in the NVSS \citep{vern19}. Recently, a depolarization study of LOFAR polarized sources found that they have a typical RM dispersion within the beam of $<0.3$~rad~m$^{-2}$ \citet{2020A&A...638A..48S}. Therefore, any radio galaxy that is embedded in an environment with RM fluctuations larger than this on scales smaller than $20"$ will not be present in the LOFAR data, but can be detected at 1.4 GHz (for RM dispersions up to a few 10s of rad~m$^{-2}$). Thus, high resolution simulations, that include radio galaxies embedded in realistic environments that have RM fluctuations on a wide range of scales, can help statistically isolate those sources whose RM variance is dominated by the local environment (e.g.~the intracluster medium) and those that are more pristine probes of cosmic filaments.

In this work we were only concerned with ideal MHD, meaning that the resistive dissipation of magnetic fields, the onset of "microscopic" plasma instabilities and other effects related to the departure from a single fluid model were not included \citep[e.g.][]{sch05}. 


Finally, in order to solely focus on the possible differences between variation of primordial seeding models, we neglected alternative scenarios for the seeding of magnetic fields related with galactic activity \citep[e.g.][]{1997ApJ...480..481K,Kronberg..1999ApJ,Volk&Atoyan..ApJ.2000,2005A&A...443..367L,donn09,2006MNRAS.370..319B,sam17}.

However, the most important differences between primordial models we find concern the very peripheral regions of galaxy clusters, or filaments and cosmic voids, in which the impact of "astrophysical" sources of magnetisation must be sub-dominant \citep[e.g.][]{va17cqg,hack19}. Therefore, the main differences found in this work are very likely to persist regardless of other sources of magnetisation in large-scales, if primordial magnetic fields do exist. 


\section{Conclusions}
\label{sec:conclusions}

We presented cosmological magneto-hydrodynamical simulations for  six different initialisations of "primordial" seed magnetic field. We adopted  models  designed to be compatible with the conservative upper limits of non-helical primordial magnetic fields derived from the latest Planck, BICEP/Keck Array, SPT  CMB anisotropy power spectra data as constrained in \cite{Paoletti:2019pdi}.
We investigated the impact of the scale dependence of primordial magnetic fields by assuming a stochastic background, spanning different possible field configurations. The scale dependence is related to the generation mechanism and advance in this topic could be important in the understanding of the physics of the early Universe. In particular, causal mechanisms, mainly related to first-order phase transitions and second-order perturbations, require a spectral index  $\alpha\geq 2.0$ that represents the extreme blue limit of our analysis. Inflationary mechanism can instead generate fields with also "red" spectra, that we sample starting from the lowest infrared limit $\alpha=-2.9$.
We have limited our analysis to the effect of primordial magnetic fields at the stochastic background level, leaving the study of a modified matter power spectrum due to magnetised cosmological perturbations for a future work.

We find that while matter halos at low-$z$ do not retain memory of seed fields (due to the fact that their magnetic field is dominated by dynamo amplification),  their outskirts, filaments and voids might retain traces of the topology and amplitude of the primordial fields. 

Our main findings can be so summarised as follows: 
\begin{itemize}
\item different models of primordial seed fields, all designed to be consistent with the most recent conservative constraints from the combination of the recent CMB data of Planck 2018 and ground based observatory BICEP/KECK and South Pole Telescope \citep{Paoletti:2019pdi}, produce significant differences in the average strength and topology of magnetic fields in the less dense cosmic environment, i.e. for $\rho \leq 10^{-30} ~ \rm g/cm^3$, even down to $z=0$;

\item depending on the amplitude of magnetic field fluctuations at small scales ($\ll 10 ~\rm Mpc$) additional gas perturbations, eventually leading to shock waves, are driven in the most rarefied cosmic gas, in addition to "standard" structure formation shocks (albeit adding little extra energy in the total volume);

\item all variations of the magnetic field scenarios have no relevant impact on the formation of structures, at least for the $\geq 10^{12} \rm M_{\odot}$ masses probed in this work;

    \item the investigated properties of magnetic fields in galaxy clusters, and of the observables connected to them, are almost invariant within $\leq ~\rm R_{\rm 200}$ of simulated clusters. Differences between models, which can be ideally detected under perfect observing conditions, are instead present in the most peripheral regions of clusters, $\geq ~\rm R_{\rm 200}$, which are at the edge of detectability with present (or future) instruments.  
    
    \item The Faraday Rotation Measure and the radio synchrotron emission produced are significantly different in the outskirts of galaxy clusters, in cosmic filaments and in voids. Differences between models can be preferentially detected in a statistical way on large cosmic volumes (in this work we tested lightcones reaching out to $z=0.8$). 
    
    \item We preliminary compared with recent available tests of the cross-correlation between galaxies and diffuse synchrotron emission using MWA \citep[][]{vern17} or of the RM structure function from physical pairs of radio sources \citep[][]{os20}. These tests suggest that existing surveys may already be employed to constrain primordial models of magnetism as some of the investigated models already produce too strong signals to be compatible with observations. The limited resolution available in these runs prevent us however from firm conclusions about which models are allowed by present radio observations.
    
    \item significant differences are in principle detectable in the average deflection angle experienced by UHECRs ($10^{20}\rm ~eV$) as they probe large scales of the seed magnetic field distributions in voids, which persists until $z=0$. 

\end{itemize}
To summarise, in this work we have established for the first time a direct link between primordial magnetic fields (as constrained by the CMB anisotropy power spectra) and the nowadays cosmic magnetism (as it can be directly observed with radio observatories and through the propagation of charged cosmic rays). As initial conditions, we considered the most conservative Planck 2018 - BICEP/KECK - SPT constraints on primordial magnetic fields, resulting from their gravitational effects on CMB anisotropies: the allowed comoving amplitude of the stochastic background ranges  
from few nano-Gauss to few pico-Gauss, depending on its scale dependence (note that tighter limits might be even obtained by considering additional effects, such as post-recombination heating, or CMB higher order statistics). The variation of the amplitude with the configuration is the key point in understanding the results: blue power spectra which increase the power on the smallest scales are already strongly constrained to very low amplitude by the CMB, whereas fields with more infrared spectra are allowed to larger amplitudes, yet with diluted power on much larger scales. The interplay between the scale dependence of the fields, and the constraints on their amplitude by the CMB, is perfectly reflected in our simulation results.

This research program is expected to  advance on different fronts, also following the deployment of new instruments.  For example future CMB experiments as LiteBIRD \citep{Hazumi:2019lys} \footnote{A joint satellite mission JAXA-US-Europe to study the large scale polarization of the CMB with unprecedented sensitivity with launch in 2027.} and ground based observatories as Simons Observatory \citep{Ade:2018sbj}\footnote{Ground based observatory that will observe the 40\% of the sky on the smallest scales with high sensitivity, will start acquiring data in 2021.}  will further tighten the constraints on the amplitude of the fields especially on very blue spectra with Simons observatory and infrared spectra with LiteBIRD reaching the subnanoGauss and subopicoGauss level \citep{Paoletti:2019pdi}. 

On the theoretical side, progresses on the impact on the ionization history \citep{Kunze:2014eka,Chluba:2015lpa,Paoletti:2018uic} could allow to further tighten the CMB constraints used here, which could be also made more accurate by including the modification to the matter power spectrum induced by primordial magnetic fields.

Jointly with the effort of increasing the resolution of the simulations, these advances can further bridge the connection between early and late cosmic magnetism to understand its origins. 

\section*{Acknowledgements}

The cosmological simulations described in this work were performed using the {\enzo} code (http://enzo-project.org). 
F.V. and S.B. acknowledge financial support from the ERC Starting Grant "MAGCOW", no.714196, and  the  usage of computational resources on Marconi at CINECA (projects INA17\_C5A38 and INA17\_C4A28 with F.V. as Principal Investigator) and at Julich Supercomputing Centre, under project STRESSICM, as well as the  usage of online storage tools kindly provided by the INAF Astronomical Archive (IA2) initiative (http://www.ia2.inaf.it). D.P. acknowledges the  usage of computational resources on Marconi and Galileo at CINECA (project INA17\_C5A42). DP and FF acknowledge financial support by ASI Grant 2016-24-H.0. MB is supported by the Deutsche Forschungsgemeinschaft (DFG, German Research Foundation) under Germany’s Excellence Strategy – EXC 2121 Quantum Universe – 390833306.\\

\section*{Data Availability}
The code used to produce the simulations discussed in this paper is public (enzo-project.org). The method to produce initial conditions necessary to perform the simulations are documented in the paper, and the 3-dimensional files can be shared by the authors upon request.

\bibliographystyle{mnras}
\bibliography{info,franco3,dan}

\appendix
\section{Resolution tests}

\begin{figure}
\includegraphics[width=0.45\textwidth,height=0.325\textwidth]{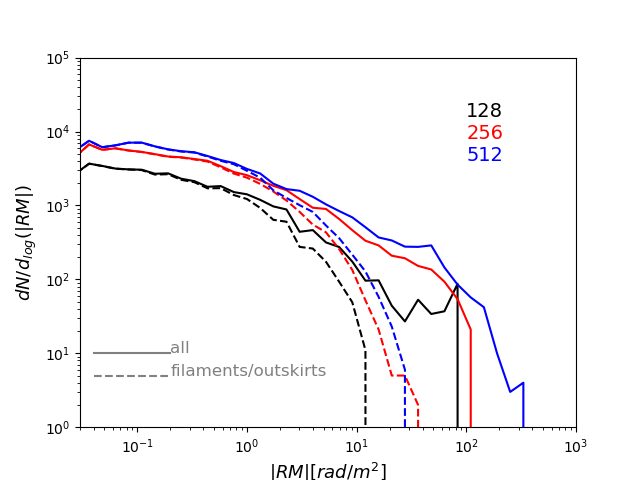}
\includegraphics[width=0.45\textwidth,height=0.325\textwidth]{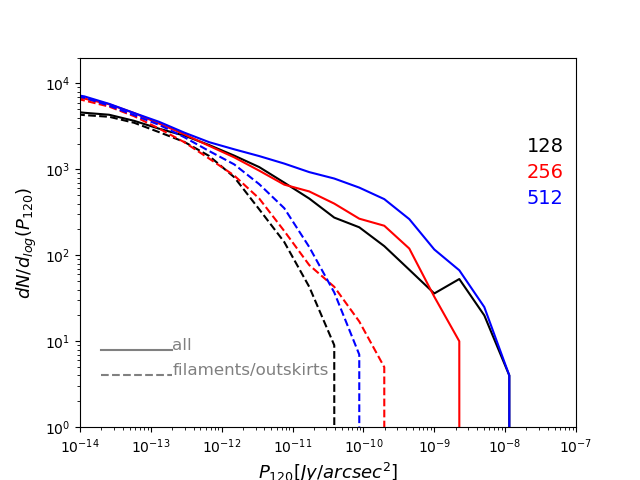}
\caption{Distribution functions of Faraday Rotation (top) and of synchrotron radio emission at 1.4 GHz (bottom) in our resolution tests, for a $1{\circ} \times 1^{\circ}$ field of view integrated for a light cone up to $z=0.8$.}
\label{fig:res2}
\end{figure}

\begin{figure}
\includegraphics[width=0.495\textwidth,height=0.325\textwidth]{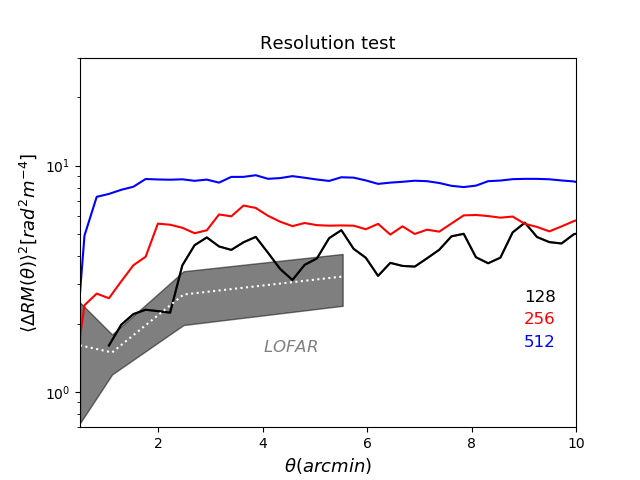}
\caption{Simulated distribution of $(\Delta RM(\Delta \theta)^2)$ as a function of angular separation for our three resolution tests for a small $25^3 \rm Mpc^3$ box.}
\label{fig:res3}
\end{figure}

To test for the dependence on the spatial/mass resolution of our results, we resimulated a smaller $25^3 ~ \rm Mpc^3$ volume with the same set of cosmological parameters of the main paper, using from $128^3$,  $256^3$ and $512^3$ cells, which gives a comoving 
spatial resolution of $195 \rm ~kpc$ (as in our main paper), $97.6 \rm ~kpc$ and $48.8 \rm ~kpc$, respectively. We limited to the simple uniform seed magnetic field model with $B_0=2 \rm ~nG$, and applied the same sub-grid model for small-scale dynamo amplification as in the main paper.  

Using the same approach of the main paper, we produced mock light-cones of radio observables up to $z=0.8$, with the limitation that the very limited volume of this box does introduce artefacts in the final light cone, as the same structures are bound to repeat several time to fill the simulated $1^\circ \times 1^\circ$ field of view. Fig.~\ref{fig:res2} shows the distribution 
of integrated Faraday Rotation and synchrotron emission for the light cones in the three runs. The distributions have a remarkably similar shape across resolutions, especially in the range of filaments and cluster outskirts, which is the most interesting for our study.

\begin{figure*}
\includegraphics[width=0.99\textwidth,height=0.3\textwidth]{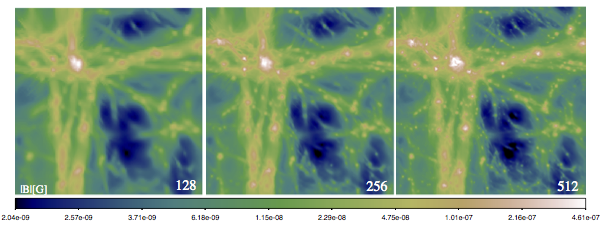}
\includegraphics[width=0.99\textwidth,height=0.3\textwidth]{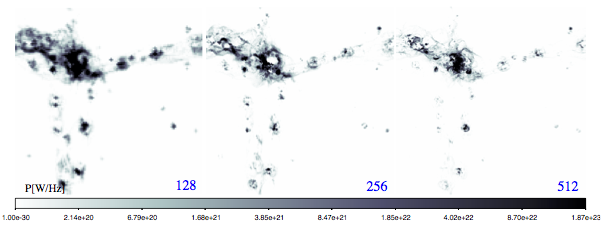}
\caption{Maps of projected magnetic field strength (top) and synchrotron radio power (bottom) for our resolution tests at $z=0$.}
\label{fig:res1}
\end{figure*}  
However, the increase of small-scale features in the gas flow internal to filaments does produce an evolution with resolution in the mock distribution of RM structure function, as shown in Fig.~\ref{fig:res3}. While the $128$ case looks in the range of what found in the main paper at the same resolution (albeit with a large scatter due to the small statistics of this box), the other two runs produced a higher level of $)\langle\Delta RM(\theta)\rangle)^2$ at most scales. Incidentally, the 256 run has a resolution close to the simulation used in our recent comparison with LOFAR data \citet{os20}, and once rescaled for the same initial magnetic field amplitude give a similar trend. 
Although not dramatic, considering the $\times 4$ increase in spatial resolution and the $\sim 4^3$ increased mass resolution, these resolution trends suggest that the absolute level of the RM structure function obtained in our main paper must be taken with care, due to the emergence of small-scale structures affects this observable more than others.

Fig.~\ref{fig:res1} gives the projected magnetic field strength across the simulated box, and the projected synchrotron radio emission from shocks, computed as in the main paper. The large scale distribution of magnetic fields is remarkably similar on all scales, which also confirms that the our sub-grid dynamo model is well behaved across resolutions, despite the progressive increase of the Reynolds number of the gas flow within galaxy clusters. On the other hand, as expected shocks are increasingly better resolved in cluster outskirts, and finer details in the morphology of synchrotron emission internal to large-scale structures appear. 

As a caveat, we shall notice that although these tests concerned the evolution of magnetic field topology/amplitude within filaments and cosmic structures, they cannot test the physical effects of sampling a higher $k_D$ (i.e. a smaller cutoff scale) in simulated primordial fields.

\end{document}